\documentclass[twocolumn,showpacs,preprintnumbers,amsmath,amssymb]{revtex4}
\usepackage{amsmath,amsfonts,latexsym,amssymb,graphics,epsfig,subfigure,color,makeidx}
\usepackage{xcolor}
\usepackage[colorlinks,
            linkcolor=blue,
            anchorcolor=blue,
            citecolor=green
            ]{hyperref}

\begin{document}

\title{Motion of test particle in rotating boson star}

\author{Yu-Peng Zhang,
	    Yan-Bo Zeng,
        Yong-Qiang Wang,
        Shao-Wen Wei,
        Yu-Xiao Liu\footnote{liuyx@lzu.edu.cn, corresponding author}
}
\affiliation{Lanzhou Center for Theoretical Physics, Key Laboratory of Theoretical Physics of Gansu Province,\\
Institute of Theoretical Physics \& Research Center of Gravitation, Lanzhou University, Lanzhou 730000, China\\
School of Physical Science and Technology, Lanzhou University, Lanzhou 730000, China}

\begin{abstract}
Motion of a test particle plays an important role in understanding the properties of a spacetime. As a new type of the strong gravity system, boson stars could mimic black holes located at the center of galaxies. Studying the motion of a test particle in the spacetime of a rotating boson star will provide the astrophysical observable effects if a boson star is located at the center of a galaxy. In this paper, we investigate the timelike geodesic of a test particle in the background of a rotating boson star with angular number $m=(1, 2, 3)$. With the change of angular number and frequency, a rotating boson star will transform from the low rotating state to the highly relativistic rapidly rotating state, the corresponding Lense-Thirring effects will be more and more significant and it should be studied in detail. By solving the four-velocity of a test particle and integrating the geodesics, we investigate the bound orbits with a zero and nonzero angular momentum. We find that a test particle can stay more longer time in the central region of a boson star when the boson star becomes from low rotating state to highly relativistic rotating state. Comparing the periapse values of the orbits obtained in rotating boson stars and the corresponding orbits of the observed stars orbiting Sagittarius $A^*$, we discuss the possible observable effects of the astrophysical boson star in the Galactic center.
\end{abstract}

\maketitle

\section{Introduction}\label{scheme1}
Black holes are the most remarkable objects predicted by general relativity, the observations of gravitational waves \cite{Abbott2016a} and black hole shadow \cite{ETH2019} provide the most strong evidence to support the existence of black holes. Besides black holes, there are still some other alternative compact objects. The globally regular solitons are a family of classical solutions of nonlinear fields with localized structures and finite energy. These objects known as the boson stars were proposed in Refs. \cite{Feinblum1968,Kaup1968,Ruffini1969}. They have been considered as important objects in the cosmology. When self-interactions of the boson fields are introduced, their properties  will be different \cite{Mielke1981,Colpi1986}. The properties of various families of spherical boson stars have been investigated \cite{Liddle1992,Lee1992,Phillippe1992}.

The constructions of rotating boson stars were first realized in Newtonian approach \cite{Silveira1995,Schupp1996}. Until 1997, the first rotating boson star in general relativity was obtained \cite{Yoshida1997a,Yoshida1997b,Yoshida1997c}. When the self-interaction of the scalar field is included, the boson star will become more massive \cite{Colpi1986}. As a different type of strong gravity system, its properties are determined by the configurations of the complex scalar field. The stationary and axisymmetric rotating boson stars are based on the scalar field with the following ansatz
\begin{equation}
\Phi=\phi(r,\theta)\exp\left[i(\omega t -m\varphi)\right],
\end{equation}
where the parameters $\omega$ and $m$ are the frequency and angular quantum number, they determine the properties of the corresponding rotating boson star. With different frequency and angular number, a boson star will have different configurations. In Ref. \cite{Kleihaus2005} the spiral-like frequency dependence of a boson star was derived and the relation between the frequency $\omega$ and the ADM mass was clarified. A rotating boson star will have ergoregion when it becomes from the low rotating state to highly relavistic rotating state, which means a rotating boson star might be unstable. The stabilities under perturbation of boson star were studied in Refs. \cite{Kleihaus2005,Cardoso2008,Kleihaus2012}. To determine the final state of an unstable rotating boson star, the 3D evolution is necessary. In Ref. \cite{Balakrishna2006}, the non-linear evolution of a spherical boson star was investigated and it was shown that an unstable boson star could collapse into a black hole. With the help of numerical relativity, the formation and stability of rotating boson stars were investigated in Ref. \cite{Sanchis2019}.

Besides the boson star constructed by a scalar field with single state, the rotating multi-state boson star was also introduced in Refs. \cite{Li2020,Li2020ffy,Zeng2021oez}. Compared with the rotating boson star constructed by a scalar field with single state, the nonlinear evolution proved that a multi-state rotating boson star could be stable \cite{Sanchis-Gual:2021edp}. The collisions of binary boson star were also investigated in Refs. \cite{Balakrishna1999,Choptuik2010,Herdeiro2019c} and the corresponding gravitational waves from such systems were obtained. Recently, there is a significant break that the head-on collision of binary Proca stars \cite{Bustillo2021} could explain the GW190521 \cite{GW190521}, which provides a possible way to detect the boson stars by using the gravitational waves. In addition to the boson stars described by scalar and Proca fields, the asymptotically flat rotating Dirac star was also proposed \cite{Herdeiro2019}.

To understand the geometry of a black hole, one usually investigates the behaviors of the geodesics of a test particle. We know that a test particle can orbit around a central black hole and its motion is dependent on the geometry of the central black hole. Observations of the Keplerian stellar orbits in the Galactic center Sagittarius A* have shown that there is a high-concentrated compact object with a roughly mass of $4\times 10^6 M_\odot$\cite{Schodel:2002py,Gillessen:2008qv}. Therefore, monitoring the stellar orbits around the supermassive compact object in the galactic center \cite{Gillessen:2008qv} could provide an effective way to dig out the properties of the central supermassive compact object. Especially for the extreme mass ratio inspiral system described by a small celestial body inspiraling into a supermassive black hole, the corresponding orbit is closely related to the properties of gravitational waves. Since a Boson star is a strong gravity system, it is important to study whether a boson star could mimic a black hole. By comparing the power spectrum of a simple accretion disk model between a spherical boson star and Schwarzchild black hole \cite{Rueda2009}, the results shows that it is possible to find a boson star that can mimic the power spectrum of the disk of a black hole with the same mass. By studying the shadow of a boson star or a Schwarzchild black hole, the possibility that the Porca star shadow mimics the Schwarzchild black hole shadow was also discussed in Ref. \cite{Herdeiro2021lwl}. Note that a boson star can have small size with a very large mass,  and there is no horizon and singularity in such system, that is to say, the geodesics in such spacetime is complete and the novel orbits will exist in such system.

In Ref. \cite{Kesden2005}, the authors first investigated the timelike geodesics in spherical boson star and discussed the corresponding gravitational wave signature of a massive test particle inspiraling into a central boson star. Later, the orbits in a spherical boson star with a nonzero angular momentum were investigated \cite{Diemer2013}, where the novel orbits that are not present in the Schwarzchild black hole were found. The motion of a test particle in a rotating boson star was first investigated in Ref. \cite{philippe2014}, where the authors showed that a rotating boson star could possess new type of orbits. They found that the orbits could pass very close to the center and they are quite different from the orbits in a Kerr black hole. The circular geodesics and thick tori around rotating boson stars were also investigated in Ref. \cite{Meliani:2015zta}. By using the null geodesics around a rotating boson star, the shadow of a boson star at the Galactic center was investigated \cite{Vincent:2015xta}, the novel shadow and chaotic lensing around boson stars were also found in Refs. \cite{Cunha2015,Cunha:2016bjh}. The light rings and light points of rotating boson stars were first derived in Ref. \cite{philippe2017}. Inspired by the light points in a rotating boson star, the static orbits in a rotating boson stars were obtained \cite{Collodel2018}, where a massive test particle could keep static with respect to an asymptotic static observer in the background of a rotating boson star. Recently, the properties of timelike circular orbits in variety of boson stars are investigated in detail \cite{Jorge2021}.

With the change of the frequency $\omega$, a boson star will transform to the more relativistic rotating state and an ergoregion will appear \cite{philippe2014}. The scalar field configuration is unstable in spacetime with ergoregion due to the superradiant scattering \cite{Cardoso2008,Friedman1978}, the related researches about such instability should be investigated and the final state of such system could be determined by using the non-linear evolution \cite{Sanchis2019}. Actually, the rotating mini boson stars are unstable in the entire parameter space. The stability of the mini rotating stars was thoroughly analyzed in \cite{Siemonsen:2020hcg} in the relativistic and Newtonian regimes, and in \cite{Nikolaieva:2021owc,Dmitriev:2021utv} in the Newtonian regimes. Ref. \cite{Siemonsen:2020hcg} showed that there is non-axisymmetric instability in mini boson stars with azimuthal number $m=(1, 2)$ and this instability will lead a rotating boson star collapse into multiple unbound nonrotating stars or form the binary black holes. Rotating boson star can be stabilized using non-linear scalar self-interactions \cite{Siemonsen:2020hcg}. In this paper, we ignore the instability of rotating boson stars. We consider a test particle that moves in the background of a boson star from the low rotating state to the highly relativistic rapidly rotating state and investigate the Lense-Thirring effects of the rotating boson stars by comparing the behaviors of the corresponding orbits. {By setting the rotating boson star with the same mass as the supermassive black hole in the center of Sagittarius A*, we will compare the periastrons and apastrons of the stellar orbits between the backgrounds of the rotating boson star and the Sagittarius A* and discuss the possible novel observable effects for a central rotating boson star.}

This paper is organized as follows. In Sec.~\ref{scheme1}, we briefly introduce the constructions of boson stars and discuss the corresponding properties of boson stars in terms of configurations of the metric component $g_{tt}$ and scalar field. In Sec.~\ref{scheme2}, we consider a test particle that moves in the equatorial plane of a boson star. We solve the corresponding four-velocity and radial effective potential. In Sec. \ref{scheme3}, by integrating the geodesics, we derive the bound timelike geodesic orbits with zero and small angular momenta and analyze the properties of orbits in different rotating boson stars from low rotating state to the highly relativistic rotating state. To observe the possible hints for distinguishing the black hole or rotating boson star, we still compare the metric functions of black hole and rotating boson star that have the same mass and spin angular momentum. The relations between the orbits obtained from the background of rotating boson star and the observed stellar orbits around Sagittarius A* \cite{Gillessen:2008qv} are still discussed. Finally, a brief conclusion and outlook are given in Sec.~\ref{Conclusion}.

\section{The Boson stars}{\label{scheme1}}

A boson star is the strong gravity system constructed from the self-graviting complex scalar field. It is described by the following action
\begin{equation}
S =  \int{d^4 x \sqrt{-g}
  \left[\frac{1}{16\pi G}R - \nabla_\mu\Phi\nabla^\mu\Phi^* - V(\Phi)\right]}.
\label{action}
\end{equation}
In this paper, the Newton gravitational constant $G$, the speed of light $c$, and the Planck constant $\hbar$ are set to be unity ($G=c=\hbar=1$), and the potential $V(\Phi)$ is taken as the simplest form
\begin{equation}
V(\Phi)=\frac{\mu^2}{\hbar^2}\Phi\Phi^*,
\label{potentialofphi}
\end{equation}
where the constant $\mu$ is the mass parameter of the scalar field $\Phi$. The boson star contracted from the potential \eqref{potentialofphi} is called the mini boson star \cite{Schunck2003}.

To get a stationary and axisymmetric rotating boson star, we take the ansatz for the scalar field $\Phi$ as
\begin{equation}
\Phi=\phi(r,\theta)\exp\left[i(\omega t -m\varphi)\right],
\label{scalarf}
\end{equation}
where $(t, r, \theta, \varphi)$ are the coordinates of the spacetime. The parameter $\omega$ is the frequency of the scalar field. The stationary and axisymmetry mean that the spacetime can possess a timelike killing vector $\xi^\mu=(\partial_t)^\mu$ and a spacelike killing vector $\eta^\mu=(\partial_\varphi)^\mu$. It is the reason that the integer $m$ in Eq. \eqref{scalarf} is the rotational angular quantum number. With the two killing vectors $\xi^\mu$ and $\eta^\mu$, we take the ansatz for the metric of the rotating boson star as follows \cite{Herdeiro2015}
\begin{eqnarray}
ds^2&=&g_{tt}dt^2+g_{rr}dr^2+g_{\theta\theta}d\theta^2+2g_{t\varphi}dt d\varphi+g_{\varphi\varphi}d\varphi^2\nonumber\\
&=&e^{2F_1}\left(\frac{dr^2}{N^2}+r^2d\theta^2\right)- e^{2F_0}Ndt^2\nonumber\\
&& +~e^{2F_2}r^2\sin^2\theta(d\varphi-Wdt)^2,
\label{metric}
\end{eqnarray}
where the lapse function $N=1$. Note that for a hairy black hole solution, the lapse function is given by
$N=1-\frac{r_h}{r}$, where $r_h$ is the horizon radius.

We have specified the symmetry of the geometry of the rotating boson star, therefore $F_0$, $F_1$, $F_2$, and $W$ should be the functions of $(r,\theta)$. The resulted field equations based on the action \eqref{action} are
\begin{eqnarray}
R_{\mu\nu}-\frac{1}{2}Rg_{\mu\nu}&=& 8\pi T_{\mu\nu},\label{greq}\\
\nabla_\mu\nabla^\mu \Phi        &=& \mu^2\Phi.\label{eqsf}
\end{eqnarray}
With the help of the metric ansatz \eqref{metric} and the field configuration \eqref{scalarf}, one can obtain the explicit forms of the field equations \eqref{greq} and \eqref{eqsf}, see the details in Ref. \cite{Herdeiro2015}. With the explicit equations of motion, one can numerically solve the differential equations to get the solutions of the rotating boson star.

The symmetry of the spacetime demands that the metric functions and scalar field should have a reflection symmetry. Thus, we should introduce the following boundary conditions at $\theta = \pi/2$:
\begin{eqnarray}
\partial_\theta F_i(r,\frac{\pi}{2})=\partial_\theta W(r,\frac{\pi}{2})=0
\label{boundary1}
\end{eqnarray}
and
\begin{eqnarray}
\left\{
    \begin{array}{cc}
    \partial_\theta \phi(r,\frac{\pi}{2})=0,~~~\textrm{even parity},\label{ds}\\
    ~~\phi(r,\frac{\pi}{2})=0,~~~\textrm{odd parity}
    \label{Ads}.
    \end{array}\right.
\label{boundary1}
\end{eqnarray}
The solution of a boson star must be regular through the whole spacetime. We set the metric functions and scalar field on the pole points at $\theta=0$ and $\theta=\pi$ as
\begin{eqnarray}
\partial_\theta F_i(r,0)\!\!&=&\!\!\partial_\theta W(r,0) \,= \phi(r,0) =0, \label{boundary2}\\
\partial_\theta F_i(r,\pi)\!\!&=& \!\! \partial_\theta W(r,\pi)= \phi(r,\pi) =0. \label{boundary3}
\end{eqnarray}
The asymptotic flat for a rotating boson star also demands that
\begin{eqnarray}
\lim_{r\to\infty} F_i = \lim_{r\to\infty} W = \lim_{r\to\infty} \phi =0.
\label{boundary4}
\end{eqnarray}
Then a rotating boson star could be obtained by numerically solving the equations \eqref{greq} and \eqref{eqsf} with the boundary conditions \eqref{boundary1}-\eqref{boundary4}. We use the same method in Ref. \cite{wang2019} to get the numerical solutions of the rotating boson star. One can also use the spectral solver KADATH \cite{philippe2010,philippe2014} to derive the solutions of the boson star. We compare our numerical results with one case from KADATH, and we find that the two solutions are perfectly consistent.

Next, we will give a brief discussion about the properties of a rotating boson star. Note that the properties of a rotating boson star are mainly controlled by the frequency parameter $\omega$ and the angular quantum number $m$. Therefore, the scalar field will have different configurations with different $\omega$ and $m$.

\begin{figure*}[!htb]
\includegraphics[width=\linewidth]{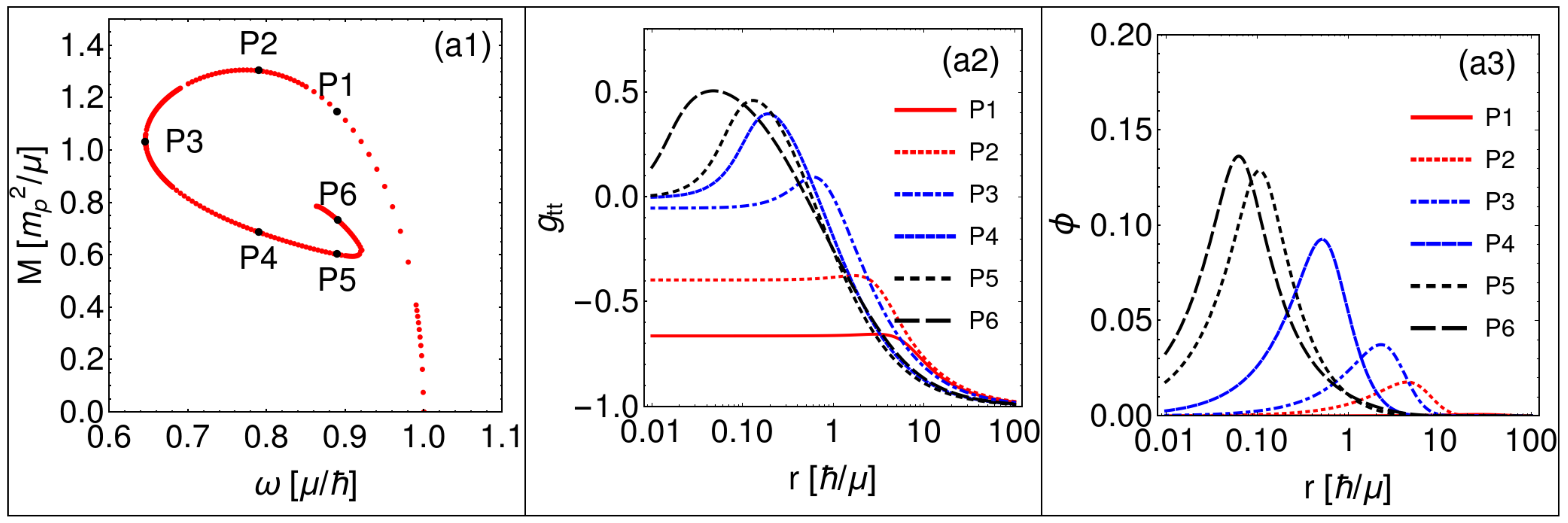}
\includegraphics[width=\linewidth]{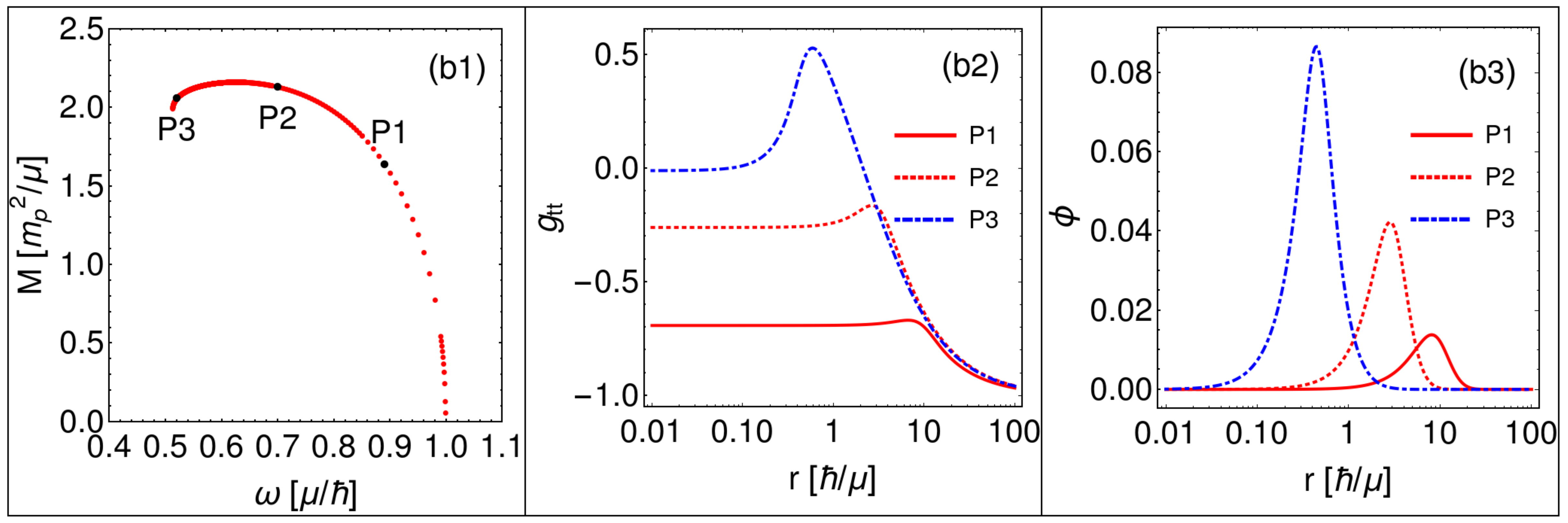}
\includegraphics[width=\linewidth]{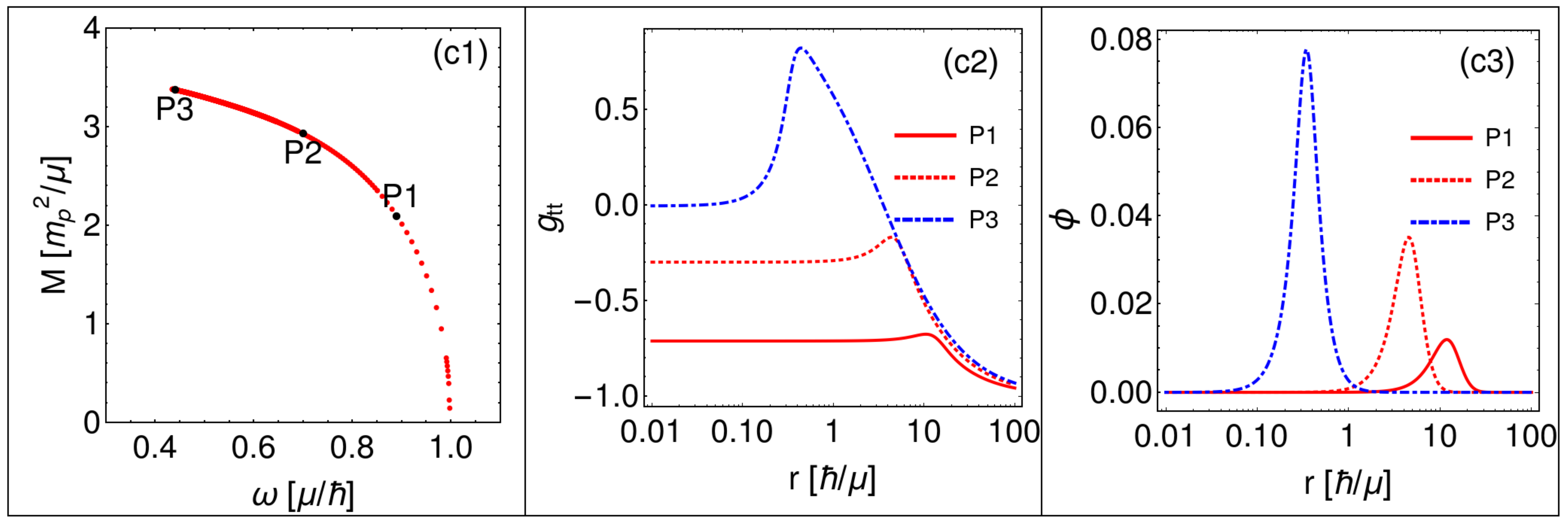}
\caption{Plots of the ADM mass $M$, metric function $g_{tt}$, scalar field modulus $\phi$ of the boson star. We set $m=1,2,3$ for subfigures (a1-a3), (b1-b3), (c1-c3), respectively. }
\label{frequency_gtt_phi}
\end{figure*}

In this paper, we only consider the rotating boson stars with angular quantum number $m=(1, 2, 3)$, the corresponding results about the ADM mass as a function of $\omega$ are given in Fig.~\ref{frequency_gtt_phi} \cite{Herdeiro2015}. A highly relativistic rapidly rotating boson star will have an ergoregion \cite{philippe2014}, one can get the ergoregion by using
\begin{equation}
g_{tt}=r^2 \sin ^2(\theta) W(r,\theta)^2 e^{2 F_2(r,\theta)}-e^{ 2 F_0(r,\theta)}>0.
\end{equation}
To study the Lense-Thirring effect of the rotating boson stars from low rotating state to highly relativisitc rotating state, we pick several solutions described by the points in Fig.~\ref{frequency_gtt_phi}. We list the corresponding frequency $\omega$, ADM mass $M$, and the dimensionless reduced angular momentum $J/M^2$ in Table \ref{mass_spin_each_points}. The corresponding profiles of the metric component $g_{tt}(r,\pi/2)$ and scalar field $\phi(r,\pi/2)$ for each point are still given in Fig.~\ref{frequency_gtt_phi}. We use the radial maximum position of the scalar field $\phi$ to define the compactness of the boson star. It is easy to see that a rotating boson star becomes more and more compact along the curve \cite{Herdeiro2015}. Besides the definition of the compactness in terms of radial maximum position of the scalar field $\phi$, one can use the perimeteral radius $r_{99}$ containing $99\%$ of the boson star mass \cite{Herdeiro2015} to get a more accurate definition. Note that, the perimeteral radius $r_{99}$ does not increase monotonically along the curve.

\begin{widetext}
	\begin{table*}[!htb]
		\begin{center}
			\caption{Parameters of boson stars considered in this paper of the points labeled in Fig. \ref{frequency_gtt_phi}.}
			\begin{tabular}{ c | c | c | c | c | c | c | c | c | c}
				\hline
				\hline
		        ~~~~model~~~~&~~~~$\omega$~~~~&~~$M/(m_p^2/\mu)$~~&~~~~$J$~~~~&~~$J/M^2$~~&~~~~model~~~~&~~$\omega$~~&~~$M/(m_p^2/\mu)$~~&~~~~$J$~~~~&~~~$J/M^2$~~~ \\
				\hline
				 $P_1 (m=1)$ & 0.8900       &  1.1524   &   1.1902  &   0.8962  & $P_1 (m=2)$ &  0.8900  &   1.6660  &   3.4484  &   1.2424     \\
				 $P_2 (m=1)$ & 0.7900       &  1.3124   &   1.3776  &   0.7998  & $P_2 (m=2)$ &  0.7000  &   2.1831  &   4.7134  &   0.9890     \\
				 $P_3 (m=1)$ & 0.6457       &  1.0309   &   0.9582  &   0.9016  & $P_3 (m=2)$ &  0.5200  &   2.1038  &   4.4003  &   0.9942     \\
				\hline
				 $P_4 (m=1)$ & 0.7900       &  0.6826   &   0.4523  &   0.9707  & $P_1 (m=3)$ &  0.8900  &   2.1543  & 6.6967    &   1.4429     \\
				 $P_5 (m=1)$ & 0.8900       &  0.5795   &   0.3491  &   0.9779  & $P_2 (m=3)$ &  0.7000  &   3.0562  & 10.0478   &   1.0757     \\
				 $P_6 (m=1)$ & 0.8910       &  0.7276   &   0.7922  &   0.9297  & $P_3 (m=3)$ &  0.4400  &   3.5577  & 12.6624   &   1.0004     \\
				\hline
				\hline
			\end{tabular}
			\label{mass_spin_each_points}
		\end{center}
	\end{table*}
\end{widetext}

\section{Velocities and effective potential}\label{scheme2}

Studying the geodesics in a given spacetime geometry is very useful to understand the properties of the spacetime, and the information of the geodesics will provide the astrophysical observable effects. In this section, we will investigate the motion of a test particle in the background of a rotating boson star. By using the effective potential method, it is easy to derive the orbits around axisymmetric and stationary objects. For simplicity, we only consider the orbits in the equatorial plane. The four-velocity is given by
\begin{equation}
u^{\mu}=\left(\frac{dt}{d\tau},\frac{dr}{d\tau},0,\frac{d\varphi}{d\tau}\right)=\left(u^t,u^r,0,u^\varphi\right),
\end{equation}
where $\tau$ is the proper time of the test particle.

The existence of the two independent Killing vectors $\xi^\mu=(\partial_t)^\mu$ and $\eta^\mu=(\partial_\varphi)^\mu$ leads to two conserved quantities for a test particle, i.e., the energy $\bar{E}$ and orbital angular momentum $\bar{J}$ of the particle per units mass:
\begin{eqnarray}
-\bar{E}&=&(\partial_t)^\mu u_{\mu}=g_{tt}u^t+g_{t\varphi}u^\varphi
\label{energy} \\
\bar{J}&=&(\partial_\varphi)^\mu u_{\mu}=g_{t\varphi}u^t+g_{\varphi\varphi}u^\varphi.
\label{angularmomentum}
\end{eqnarray}
We set the mass of the test particle be one. By solving Eqs. (15) and (16), we get the components  $u^t$ and $u^\varphi$ as follows
\begin{eqnarray}
u^t &=&\frac{\bar{E} g_{\varphi\varphi} + \bar{J} g_{t\varphi}}{g_{t\varphi}^2 - g_{tt}g_{\varphi\varphi}},\label{ut_eq}
\\
u^\varphi &=& \frac{\bar{E} g_{t\varphi} + \bar{J} g_{tt}}{g_{tt}g_{\varphi\varphi}- g_{t\varphi}^2}.\label{uphi_eq}
\end{eqnarray}
Then, by using the relation $u^{\mu} u_{\mu}= -\varsigma^2$ with $\varsigma^2=c^2$ and $\varsigma^2=0$ for timelike and null geodesics, respectively, one can give the radial component of the four-velocity $u^r$
\begin{equation}
\left(u^r\right)^2=-\frac{\varsigma^2+g_{\varphi\varphi}u^\varphi u^\varphi + 2 g_{t\varphi}u^t u^\phi + g_{tt}u^t u^t}{g_{rr}}.
\label{velocityur}
\end{equation}
In this paper, we only consider the timelike geodesics.

It is easy to determine the radial motion of the test particle by using the effective potential. To obtain the effective potential, we decompose the form of Eq. \eqref{velocityur} as follows
    \begin{eqnarray}
    (u^r)^2 &=& \left(A \bar{E}^2+B \bar{E}+C\right) \nonumber \\
    &=& A \left(\bar{E}-\bar{E}_{+}\right)
     \left(\bar{E}-\bar{E}_{-}\right),
    \label{effectivepotentiala}
    \end{eqnarray}
where the functions $A$, $B$, $C$, $\bar{E}_{+}$, and $\bar{E}_{-}$ are given by
\begin{eqnarray}
A&=&\frac{g_{\varphi\varphi}}{g_{rr}\left(g_{t\varphi}^2 - g_{\varphi\varphi}g_{tt}\right)},\\
B&=&\frac{2 \bar{J} g_{t\varphi}}{g_{rr}\left(g_{t\varphi}^2 - g_{\varphi\varphi}g_{tt}\right)},\\
C&=&\frac{-c^2 g_{t\varphi}^2 + \left(\bar{J}^2 + c^2 g_{\varphi\varphi}\right)g_{tt}}{g_{rr}\left(g_{t\varphi}^2 - g_{\varphi\varphi}g_{tt}\right)},\\
\bar{E}_{+} &=& \frac{-B+\sqrt{B^2-4AC}}{2A},\\
\bar{E}_{-} &=& \frac{-B-\sqrt{B^2-4AC}}{2A}.
\end{eqnarray}
The effective potential of a test particle is defined by the positive square root of Eq. (\ref{effectivepotentiala})
\begin{equation}
V_{\text{eff}}  = \bar{E}_{+}.
\label{effective-potential-r}
\end{equation}
The positive square root corresponds to the four-momentum pointing toward future, while the negative one
\begin{equation}
V_{\text{eff}}'  = \bar{E}_{-}
\label{effective-potential-r-negative}
\end{equation}
corresponds to the past-pointing four-momentum \cite{misnerthornewheeler}. By using Eq. \eqref{uphi_eq}, an angular potential is introduced \cite{Collodel2018} as follows
\begin{equation}
u^{\varphi}=\frac{g_{t\varphi}}{g_{tt}g_{\varphi\varphi} - g_{t\varphi}^2}
   \left(\bar{E}-V_{\varphi}\right),
\end{equation}
where
\begin{equation}
V_{\varphi}=-\frac{\bar{J} g_{tt}}{g_{t\varphi}}.
\label{effective-potential-phi}
\end{equation}

For a test particle with a fixed energy and angular momentum, its motion could be determined by using the effective potentials \eqref{effective-potential-r-negative} and \eqref{effective-potential-phi}.
From the metric \eqref{metric}, we have
\begin{eqnarray}
g_{tt}(r,\theta) &=& -e^{2F_0}+e^{2F_2}r^2\sin^2(\theta)W^2,\\
g_{rr}(r,\theta) &=&  e^{2F_1},\\
g_{t\varphi}(r,\theta) &=& -r^2e^{2F_2}\sin^2(\theta)W,\\
g_{\varphi\varphi}(r,\theta) &=& e^{2F_2}r^2\sin^2(\theta).
\end{eqnarray}
We have obtained the numerical solutions of the metric functions, from which the effective potentials \eqref{effective-potential-r} and \eqref{effective-potential-phi} could be derived. Next, we will investigate the timelike geodesics for rotating boson stars.

\begin{figure*}[!htb]
\includegraphics[width=\linewidth]{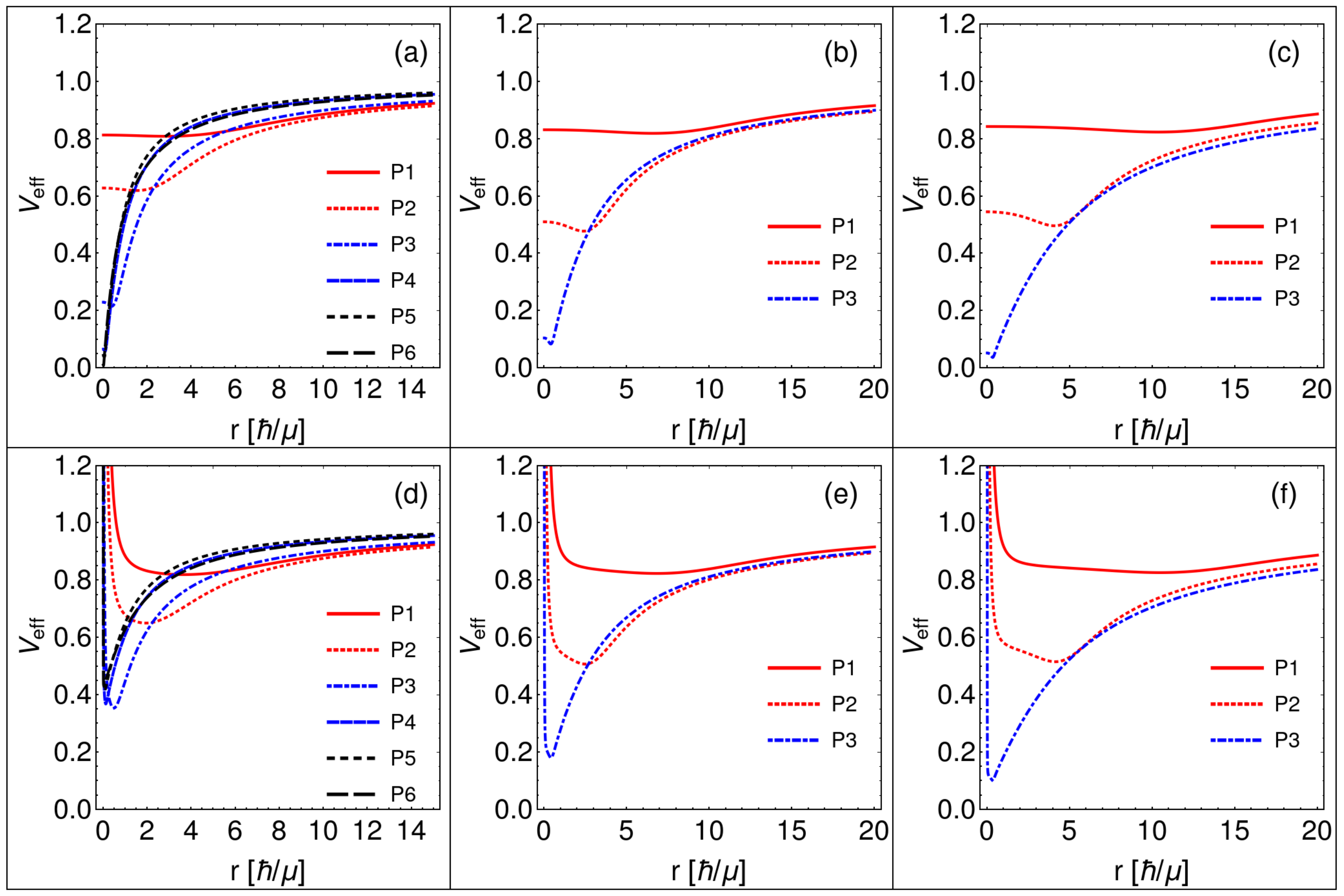}
\caption{Profiles of the effective potential (\ref{effective-potential-r}) of a test particle in the boson star with $\bar{J}=0$ (up channel) and $\bar{J}=0.5$ (down channel). The results in subfigures (a) and (d) are for the boson star with $m=1$. The results in subfigures (b) and (e) are for the boson star with $m=2$. The results in subfigures (c) and (f) are for the boson star with $m=3$.}
\label{eff_potential_m123}
\end{figure*}

\begin{figure*}[!htb]
\includegraphics[width=\linewidth]{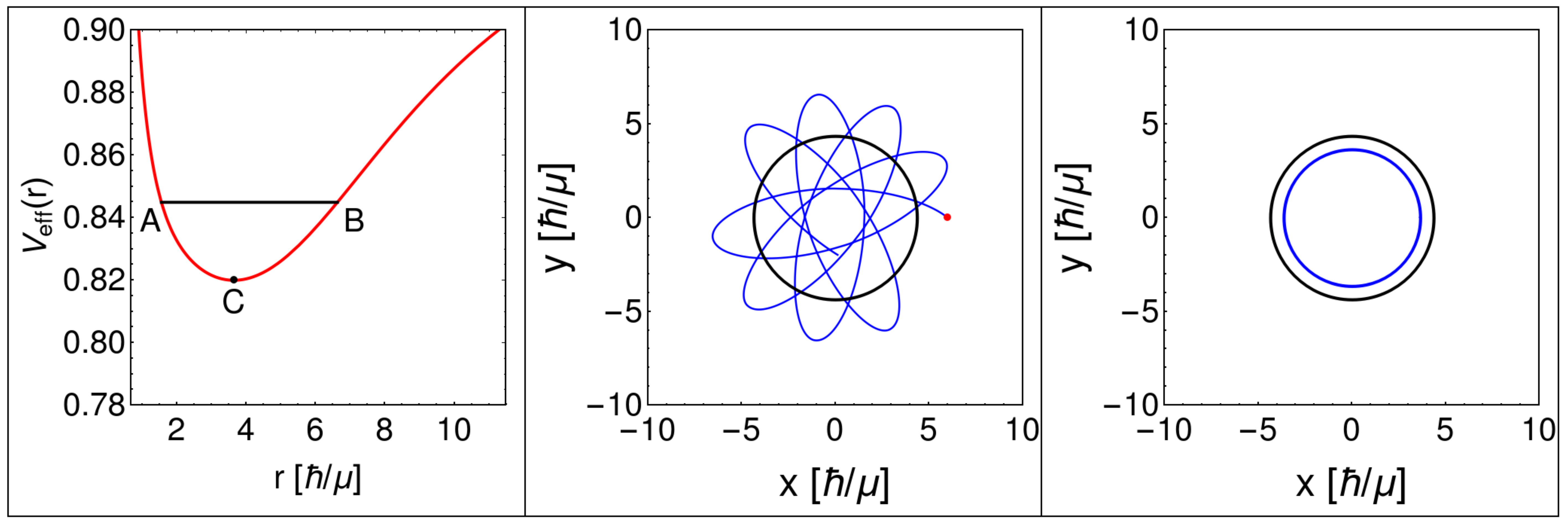}
\caption{The effective potential (\ref{effective-potential-r}) and orbit of a test particle with angular momentum $\bar{J}=0.5$ in the case of the boson star with $m=1$ and $\omega=0.79$.  The horizontal line in the left figure denotes the energy of the test particle, which is $\bar{E}=0.84493$. The middle figure describes the orbit of the test particle with energy  $\bar{E}=0.84493$. The right figure is the circular orbit with energy  $\bar{E}=0.81998$. The black circles in the middle and right figures are the radial position of the maximum scalar field. The orbit of the test particle starts from the red point. The meanings of red points and black circles in the following figures are the same as in this figure. We set the coordinates $x=r\sin\varphi$ and $y=r\cos\varphi$, which are the same in the following figures.}
\label{eff_potential_orbits_079}
\end{figure*}

Obviously, the effective potentials are dependent on the angular momentum $\bar{J}$. By specifying the energy $\bar{E}$ and angular momentum $\bar{J}$ of a test particle, we can get the orbits of a test particle in a rotating boson star. To deepen our understanding of the effective potential \eqref{effective-potential-r}, we give the plots to describe it as a function of $r$ in Fig.~\ref{eff_potential_m123}. Note that, the energy of a test particle is constant along the geodesic if we ignore the gravitational radiation. Therefore, when the angular momentum and energy are fixed, the motion status of a test particle is determined. For example, when the energy of a test particle $\bar{E}=V_{\text{eff}}(r_a)$, the allowed radial integral of the test particle is described by the horizontal line in Fig.~\ref{eff_potential_orbits_079}. When the energy of a test particle is equal to the minimum of the corresponding effective potential, the test particle will be fixed in a stable circular orbit.

We can see that the values of the effective potentials with a zero angular momentum at origin in Fig.~ \ref{eff_potential_m123} are finite. When the energy of the test particle is larger than the value of the effective potential at origin, it could move through the center of the boson star. For the orbits with a nonzero angular momentum in Fig.~\ref{eff_potential_m123}, the value of the effective potential at origin will approach to infinity, which means the test particle can not pass through the center of the boson star.

\section{Orbits of the test particle}\label{scheme3}

The significant difference between a boson star and a black hole is that a boson star does not have a singularity and event horizon, which means the geodesics in a boson star are complete and a test particle might move in the whole background of a boson star. If we consider a boson star as an alternative model of a black hole, the trajectory of a small star moving in a supermassive boson star will be different and this system will have novel astrophysical observable effects.

\begin{figure*}[!htb]
\includegraphics[width=\linewidth]{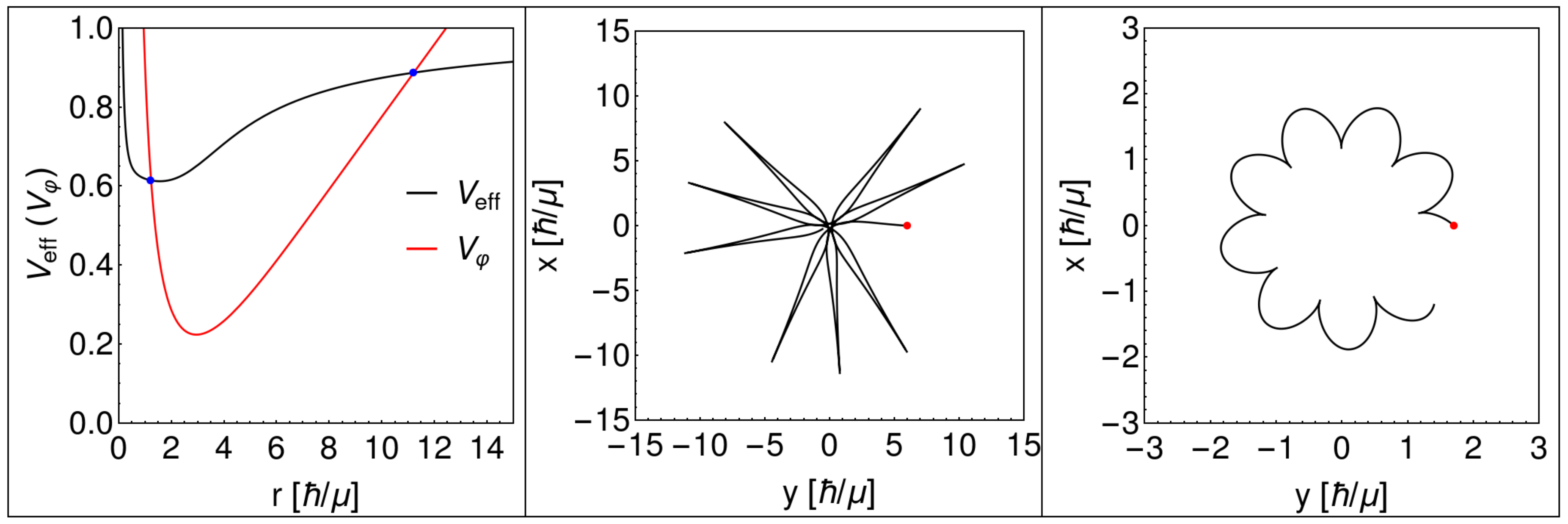}
\caption{The orbit of a test particle with angular momentum $\bar{J}=-0.248$ in the rotating boson star with $m=1$ and $\omega=0.79$. The velocity of the test particle satisfies $u^r=0$ and $u^\varphi=0$ in the peaks of the orbit.}
\label{orbits_static}
\end{figure*}

In Refs.~\cite{philippe2014,Grould12017,Collodel2018}, the authors have found some novel geodesic orbits that a black hole does not have. When a test particle moves along these orbits, the radial and angular velocities at some special points will be zero, i.e., $u^r=u^\varphi=0$. Here, we give a simple summary about the  orbits found in  Refs.~\cite{philippe2014,Grould12017,Collodel2018}. When the energy $\bar{E}$ and angular momentum $\bar{J}$ satisfy
\begin{equation}
\bar{E}-V_{\varphi}=\bar{E}+\frac{ \bar{J} g_{tt}} {g_{t\varphi}}=0
\end{equation}
and
\begin{equation}
\bar{E}-V_{\text{eff}}=0.
\end{equation}
The test particle could be initially rest \cite{philippe2014} or always rest at a special position of the boson star~\cite{Collodel2018}. Here, we give the effective potentials $V_{\text{eff}}$ and $V_\varphi$ and the corresponding orbits in Fig.~\ref{orbits_static} for the test particle with angular momentum $\bar{J}=-0.248$ in the rotating boson star with $\omega=0.79$ and $m=1$. The same kind of orbits were also found in Refs. \cite{philippe2014,Grould12017}.

We have given the profiles of the metric functions and scalar field of the boson star with different frequency $\omega$ and angular number $m$ in Fig.~\ref{frequency_gtt_phi}. With the  change of the frequency $\omega$, the compactness of the boson star will be different and some new orbits will appear. In this paper, we focus on the motion of a test particle in the background of the boson star from the slowly rotating state to the highly relativistic rapidly rotating state. The angular number $m$ is taken as $m=(1, 2, 3)$.

Based on the effective potentials, one can obtain the orbit of a test particle when the energy $\bar{E}$ and angular momentum $\bar{J}$ are fixed. When a test particle is moving along the bound orbit, the radial position of this particle will be in a finite range, for example, the orbit in Fig.~\ref{eff_potential_orbits_079} is a bound orbit. In this paper, we focus on the bound orbits of a test particle in the rotating boson star. We consider the solutions of the boson star described by the points in Fig.~\ref{frequency_gtt_phi}. The energies $\bar{E}$ of the test particle with bound orbits in all backgrounds are given in Table~\ref{energy_angular_momentum_each_points}.

\begin{widetext}
\begin{table*}[!htb]
\begin{center}
\caption{Energy of the test particle with the bound orbits in different boson star solutions.}
\begin{tabular}{ c |c | c|c |c |c|c |c}
\hline
\hline
~~~~$\bar{J}$~~~~&~~~~$m$~~~~&~~~~~~~$P_1$~~~~~~~&~~~~~~~$P_2$~~~~~~~&~~~~~~~$P_3$~~~~~~~&~~~~~~~$P_4$~~~~~~~&~~~~~~~$P_5$~~~~~~~&~~~~~~~$P_6$~~~~~~~ \\
\hline
 0         & 1         & 0.8323  &  0.7961   &   0.8382   &  0.8923    &   0.90504  &   0.88512      \\
 0         & 2         & 0.8225  &  0.7539   &   0.7684   &  -          &   -        &   -            \\
 0         & 3         & 0.8474  &  0.8114   &   0.7884   &  -          & -          &   -            \\
\hline
 0.5        & 1         & 0.8323  &  0.7961   &   0.8382   &  0.8922    &   0.9050  &   0.8851      \\
 0.5       & 2         & 0.8263  &  0.7592   &   0.7733   &  -          &   -        &   -            \\
 0.5       & 3         & 0.8492  &  0.8134   &   0.7906   &  -          & -          &   -            \\
\hline
\hline
\end{tabular}
\label{energy_angular_momentum_each_points}
\end{center}
\end{table*}
\end{widetext}
Note that, the boson stars described by the points in Fig.~\ref{frequency_gtt_phi} may have an ergoregion. To make sure a test particle could move into the ergoregion, we should check if $(u^r)^2\geq0$. By using the data in Table~\ref{energy_angular_momentum_each_points} and the positive root and negative root of the effective potentials, we make sure that $(u^r)^2\geq0$ and a test particle could move into the ergoregion. We consider two kinds of orbits, the zero and nonzero angular momentum orbits. We take the apastrons for each orbit as follows
\begin{eqnarray}
r_{\text{ap}}/\bar{M}=\left\{
    \begin{array}{cc}
    6,~~~~\textrm{$m=1,~\bar{J}=0.0$},\\
    6,~~~~\textrm{$m=1,~\bar{J}=0.5$},\\
    8,~~~~\textrm{$m=2,~\bar{J}=0.0$},\\
    8,~~~~\textrm{$m=2,~\bar{J}=0.5$},\\
    15,~~~\textrm{$m=3,~\bar{J}=0.0$},\\
    15,~~~\textrm{$m=3,~\bar{J}=0.5$},\\
    \end{array}\right.
\label{orbitsvalues}
\end{eqnarray}
where $\bar{M}=m_p^2/\mu$.

\begin{figure*}[!htb]
	\includegraphics[width=\linewidth]{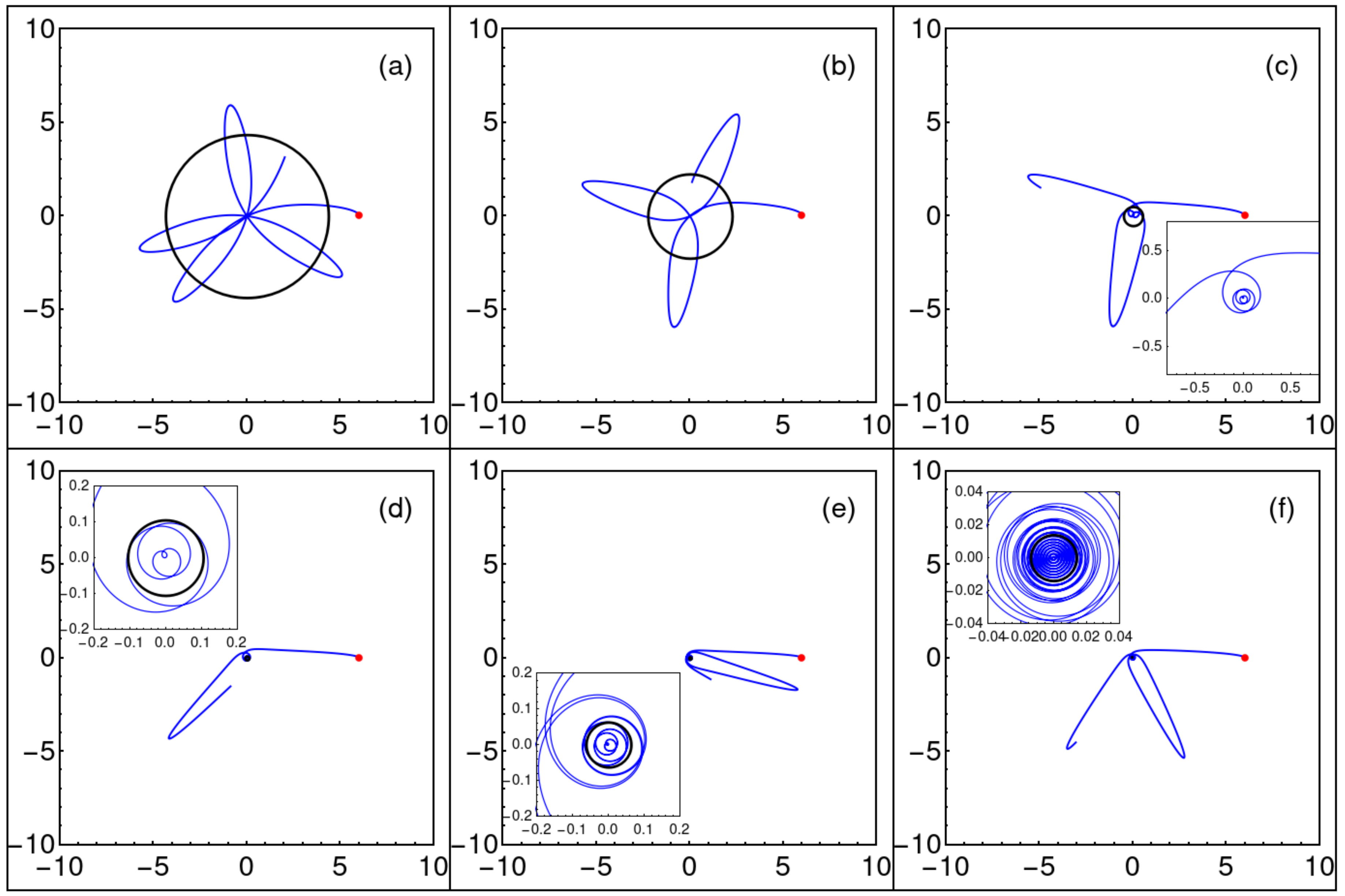}
	\caption{Zero-angular-momentum orbits of a test particle in the rotating boson star with $m=1$, where the orbits in subfigures (a), (b), (c), (d), (e), and (f) are in the background of the boson stars described respectively by the points $P_1$, $P_2$, $P_3$, $P_4$, $P_5$, and $P_6$ in subfigure (a1) of Fig.~\ref{frequency_gtt_phi}. The energies of these orbits are given in Table~\ref{energy_angular_momentum_each_points}. The inset in each subfigure of (d), (e), and (f) is a zoom of the center region in the corresponding figure.}
	\label{eff_potential_m1_j0}
\end{figure*}

\begin{figure*}[!htb]
	\includegraphics[width=\linewidth]{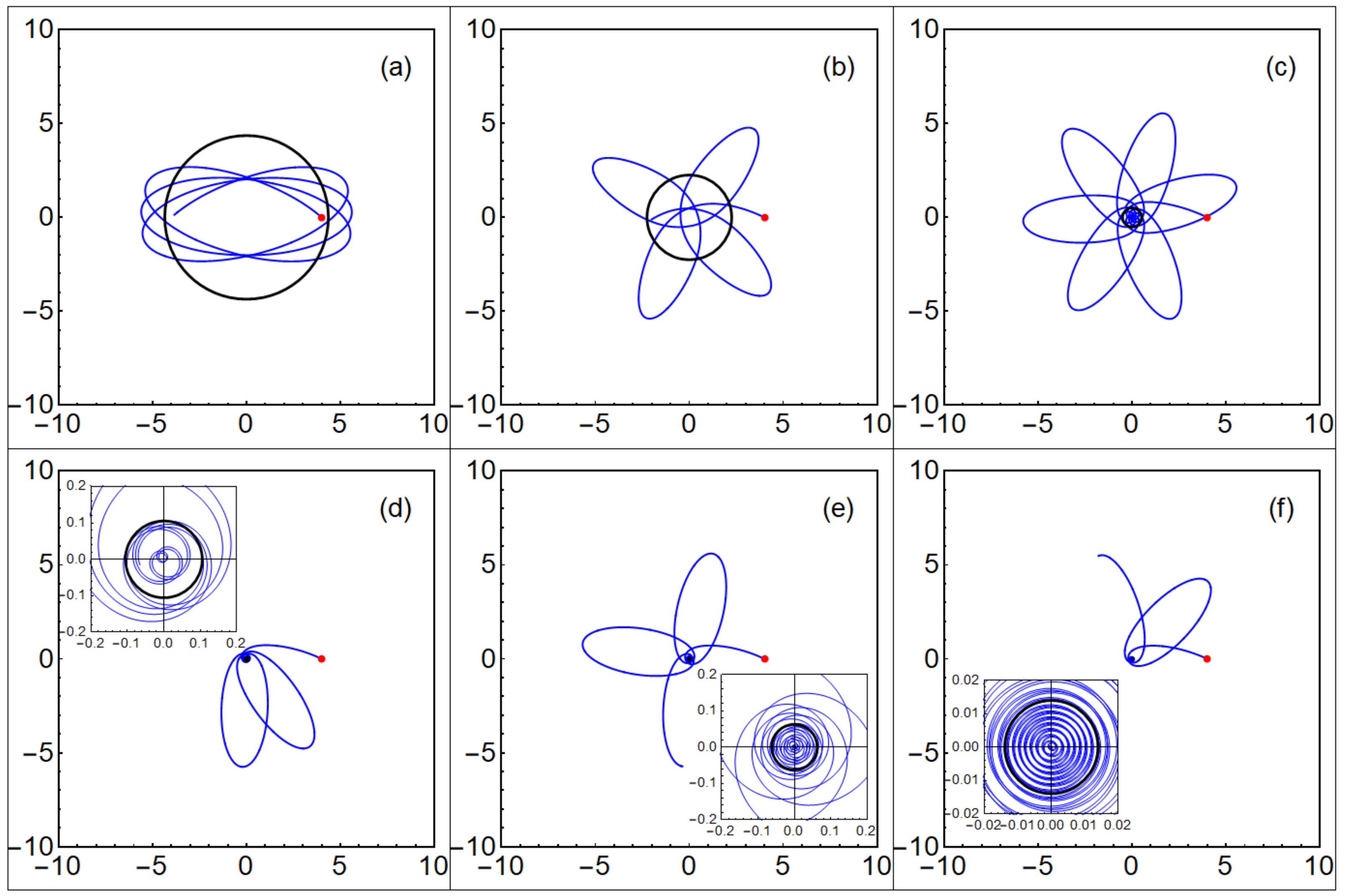}
	\caption{Orbits of a test particle with angular momentum $\bar{J}=0.5$ in the rotating boson star with $m=1$, where the setup of each figure is the same as Fig.~\ref{eff_potential_m1_j0}. The energies of these orbits are given in Table~\ref{energy_angular_momentum_each_points}. }
	\label{eff_potential_m1_j005}
\end{figure*}

Now, we have all the information to get the orbits of the test particle in the rotating boson stars. Next, we give the orbits of the test particle in the following part. The orbits of the test particle with a zero and nonzero angular momentum in the rotating boson star with $m=1$ are given in Figs.~\ref{eff_potential_m1_j0} and \ref{eff_potential_m1_j005}. We integrate the geodesics to obtain the orbits. We let our orbits have the same start point. For the zero angular momentum orbits, our choices of the energies for the orbits are larger than the values of the effective potentials at origin. Therefore, a test particle could pass through the center of a boson star. For the nonzero angular momentum orbits, a test particle can not pass through the center of the rotating boson star. With the change of the frequency $\omega$, the boson star transforms from the low rotating state to the highly relativistic rapidly rotating state. The behaviors of the angular velocity agree with this change.

\begin{figure*}[!htb]
\includegraphics[width=\linewidth]{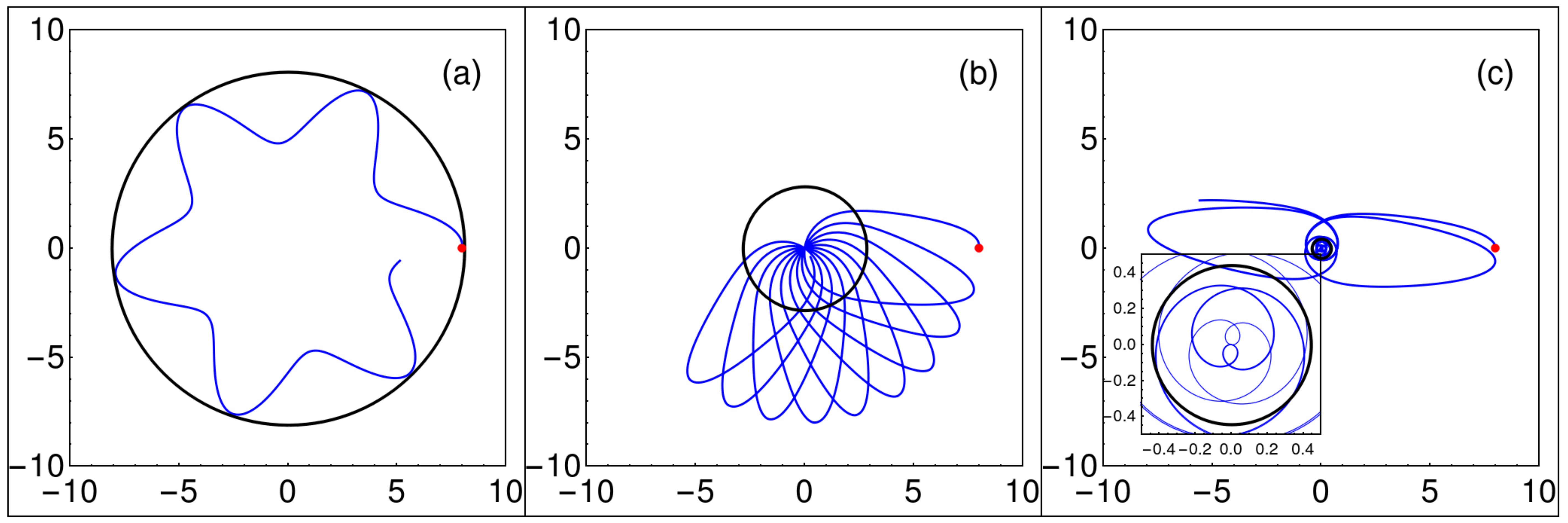}
\caption{Zero-angular-momentum orbits of a test particle in the rotating boson star with $m=2$. The orbits in subfigures (a), (b), and (c) are related with the points $P_1$, $P_2$, and $P_3$ in subfigure (b1) of Fig.~\ref{frequency_gtt_phi}. The energies of \textbf{these} orbits are given in Table~\ref{energy_angular_momentum_each_points}. }
\label{eff_potential_m2_j0}
\end{figure*}

\begin{figure*}[!htb]
\includegraphics[width=\linewidth]{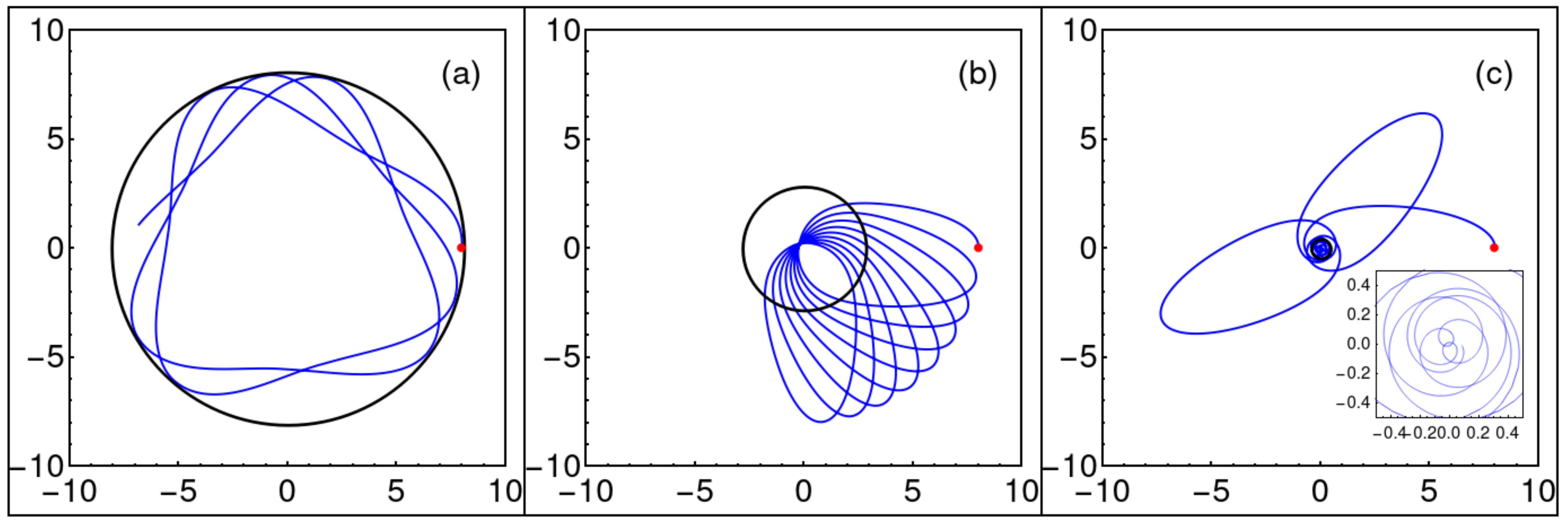}
\caption{Orbits of a test particle with angular momentum $\bar{J}=0.5$ in the rotating boson star with $m=2$, where the orbits in subfigures (a), (b), and (c) are related with the same points described in Fig.~\ref{eff_potential_m2_j0}. The energies of these orbits are given in Table~\ref{energy_angular_momentum_each_points}. }
\label{eff_potential_m2_j005}
\end{figure*}

\begin{figure*}[!htb]
\includegraphics[width=\linewidth]{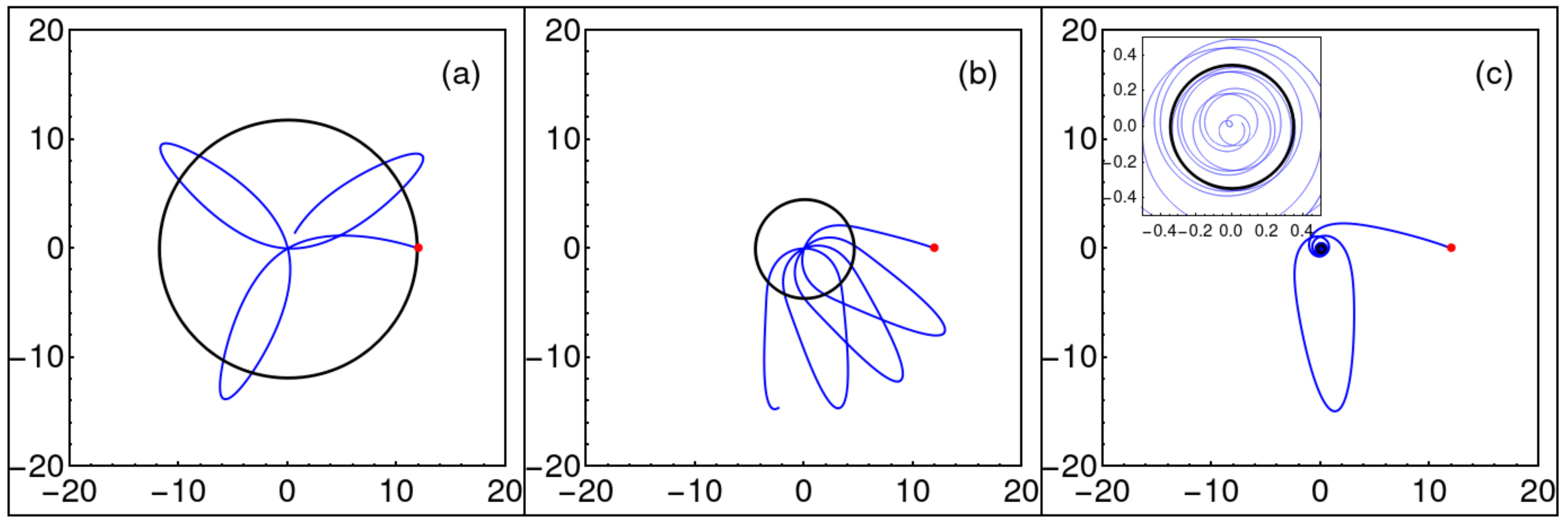}
\caption{Zero-angular-momentum orbits of a test particle in the rotating boson star with $m=3$, where the orbits in subfigures (a), (b), and (c) are related with the points $P_1$, $P_2$, and $P_3$ in subfigure (c1) of Fig.~\ref{frequency_gtt_phi}. The energies of the orbits are given in Table~\ref{energy_angular_momentum_each_points}. }
\label{eff_potential_m3_j0}
\end{figure*}

\begin{figure*}[!htb]
\includegraphics[width=\linewidth]{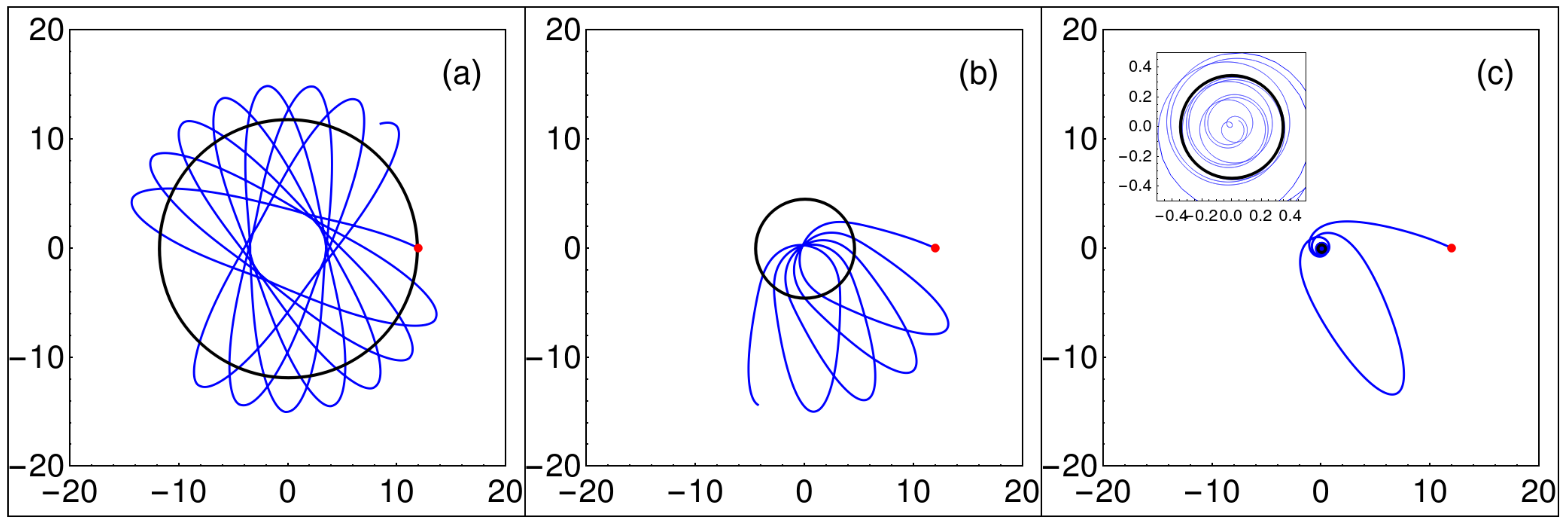}
\caption{Orbits of a test particle with angular momentum $\bar{J}=0.5$ in the rotating boson star with $m=3$. The orbits in subfigures (a), (b), and (c) are related with the points in subfigure (c1) in Fig.~\ref{frequency_gtt_phi}. The energies of the orbits are given in Table~\ref{energy_angular_momentum_each_points}. }
\label{eff_potential_m3_j005}
\end{figure*}

The orbits of a test particle with a zero and nonzero angular momentum in the rotating boson star with $m=2$ are given in Figs.~\ref{eff_potential_m2_j0} and \ref{eff_potential_m2_j005}. Here, we also let the test particle starts from the same point, we choose its energy equals to the value of the effective potential at $r=8$. With our choices about the energies of the orbits, the test particle can not pass through the center of the rotating boson star, its orbit is given in subfigure (a) of Fig.~\ref{eff_potential_m2_j0}. The orbits of the test particle with a zero and a nonzero angular momentum in the rotating boson star with $m=3$ are given in Figs.~\ref{eff_potential_m3_j0} and \ref{eff_potential_m3_j005}. With the change of the frequency $\omega$, the Lense-Thirring effects gradually become stronger and the similar behaviors to the orbits in the rotating boson star with $m=1$ also occur. The similar orbits for the low-rotating boson star were also found in Ref. \cite{Grould12017}. Due to the relations between the energy of a test particle and the value of the corresponding effective potential at center, our results include the orbits that passes or does not pass the center of the rotating boson star.

We have given the zero and nonzero angular momenta orbits of the boson stars described by the points in the subfigures (a1), (b1), and (c1) in Fig.~\ref{frequency_gtt_phi}. The periapse values and orbital eccentricities of the orbits in Figs. \ref{eff_potential_m1_j0}, \ref{eff_potential_m1_j005}, \ref{eff_potential_m2_j0}, \ref{eff_potential_m2_j005}, \ref{eff_potential_m3_j0}, and \ref{eff_potential_m3_j005} are give in Table \ref{our_orbits_periapse}. With the change of the frequency $\omega$ along the curve in subfigures (a1), (b1), and (c1) in Fig.~\ref{frequency_gtt_phi}, the boson star gradually becomes more and more compact and relativistic rapidly rotating. Especially when a test particle moves in the ergoregion of the boson star, the Lense-Thirring effects of the rotating boson star will make a significant contribution to the motion of the test particle, see the corresponding orbits in Figs.~\ref{eff_potential_m1_j0}, \ref{eff_potential_m1_j005}, \ref{eff_potential_m2_j0}, \ref{eff_potential_m2_j005}, \ref{eff_potential_m3_j0}, and \ref{eff_potential_m3_j005}. For example, when the test particle is moving along the orbit in subfigure (f) of Fig.~\ref{eff_potential_m1_j0}, it could pass through the center of the rotating boson star and will stay in the central region for a long time.

\begin{table*}[!htb]
	\begin{center}
		\caption{Values of periapse and orbital eccentricity of the orbits in Figs. \ref{eff_potential_m1_j0}, \ref{eff_potential_m1_j005}, \ref{eff_potential_m2_j0}, \ref{eff_potential_m2_j005}, \ref{eff_potential_m3_j0}, and \ref{eff_potential_m3_j005}.}
		\begin{tabular}{ c| c| c| c| c| c||c| c| c| c| c| c }
			\hline
			\hline
			~~$\bar{J}$~~&~~m~~&orbits&~~$r_a/\bar{M}$~~&~~$r_p/\bar{M}$~~&~~~$e$~~~&~~$\bar{J}$~~&~~m~~&orbits&~~~$r_a/\bar{M}$~~~&~~~$r_p/\bar{M}$~~~&~~~$e$~~~\\
			\hline
			0.0000       &   1 & a    &  6.0000 & 0.0000     &	1.0000   &0.0000      &  2  & a    &  8.0000        & 0.0000	&  1.0000    \\
			0.0000       &   1 & b    &  6.0000 & 0.0000     &	1.0000   &0.0000      & 2   & b    &  8.0000        & 0.0000    &  1.0000    \\
			0.0000       &   1 & c    &  6.0000 & 0.0000     &	1.0000   &0.0000      &	2   & c    &  8.0000        & 0.0000    &  1.0000    \\
			\cline{7-12}
			0.0000       &   1 & d    &  6.0000 & 0.0000     &	1.0000   &0.5000      &	2   & a    &  8.0000        & 5.5347    &  0.1822  \\
			0.0000       &   1 & e    &  6.0000 & 0.0000     &	1.0000   &0.5000      &	2   & b    &  8.0000        & 0.2674    &  0.9353  \\
			0.0000       &   1 & f    &  6.0000 & 0.0000     &	1.0000   &0.5000      &	2   & c    &  8.0000        & 0.0107    &  0.9973  \\
			\hline
			0.5000       &   1 & a    &  6.0000 & 2.0186    &	0.4965   &0.5000      &	3   & a   &  15.0000    & 0.0000        &  1.0000\\
			0.5000       &   1 & b    &  6.0000 & 0.4630    &	0.8567   &0.5000      &	3   & b   &  15.0000    & 0.0000        &  1.0000\\
			0.5000       &	 1 & c    &  6.0000 & 0.0512    &	0.9831   &0.5000      &	3   & c   &  15.0000    & 0.0000        &  1.0000\\
			\cline{7-12}
			0.5000       &	1  & d    &  6.0000 & 0.0066    &	0.9978   &0.5000      &	3   & a   &  15.0000     & 3.4730    &  0.6240\\
			0.5000       &	1  & e    &  6.0000 & 0.0034    &	0.9989   &0.5000      &	3   & b   &  15.0000     & 0.2701    &  0.9646\\
			0.5000       &	1  & f    &  6.0000 & 0.0003    & 0.9999     &0.5000      &	3   & c   &  15.0000     & 0.0024    &  0.9997\\
			\hline
			\hline
		\end{tabular}
		\label{our_orbits_periapse}
	\end{center}
\end{table*}

\begin{figure*}[!htb]
	\includegraphics[width=\linewidth]{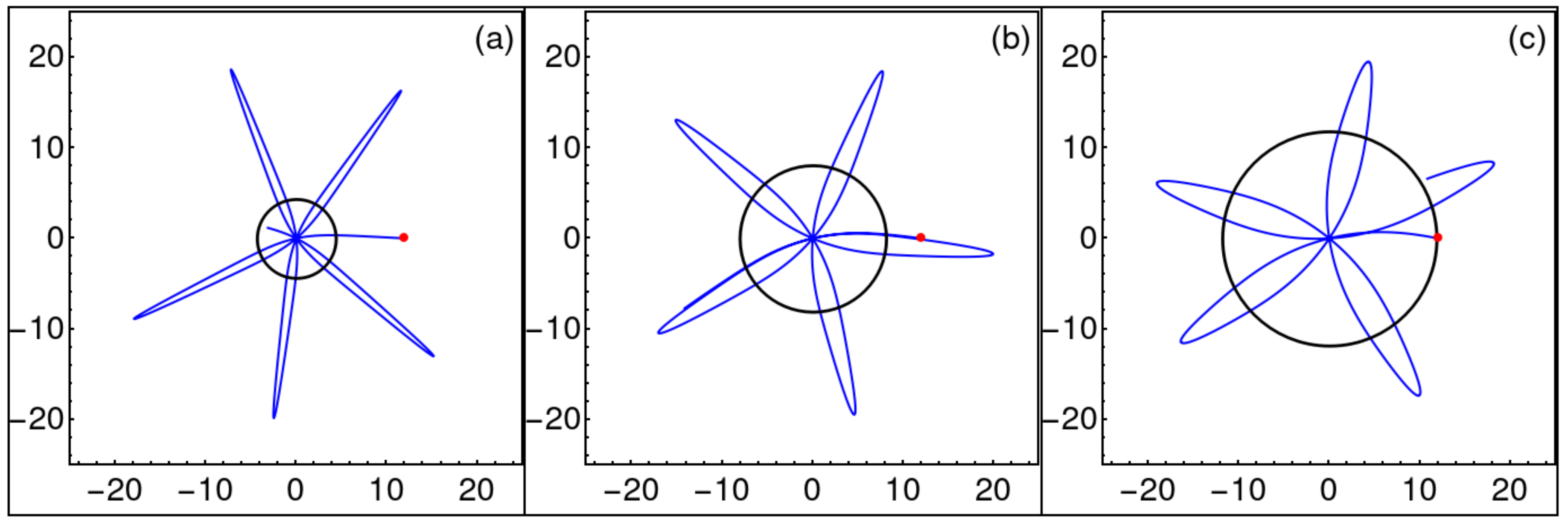}
	\caption{Zero-angular-momentum orbits of a test particle in the rotating boson star with frequency $\omega=0.89$. The orbits in subfigures (a), (b), (c) are in the backgrounds of the rotating boson star with $m=1$, $m=2$, and $m=3$, respectively. }
	\label{orbits_m123_j0}
\end{figure*}

To compare orbits in the boson star with same frequency $\omega$ but different angular number $m$, we also give the corresponding orbits of a test particle with a zero angular momentum in Fig.~\ref{orbits_m123_j0}. These orbits are set to have the same maximum radius $r_{\text{max}}=20$. It can be seen that the Lense-Thirring effect on the orbits increases with the angular number $m$. This behavior is the same as the result in Ref. \cite{philippe2014}.

\begin{figure}[!htb]
\includegraphics[width=\linewidth]{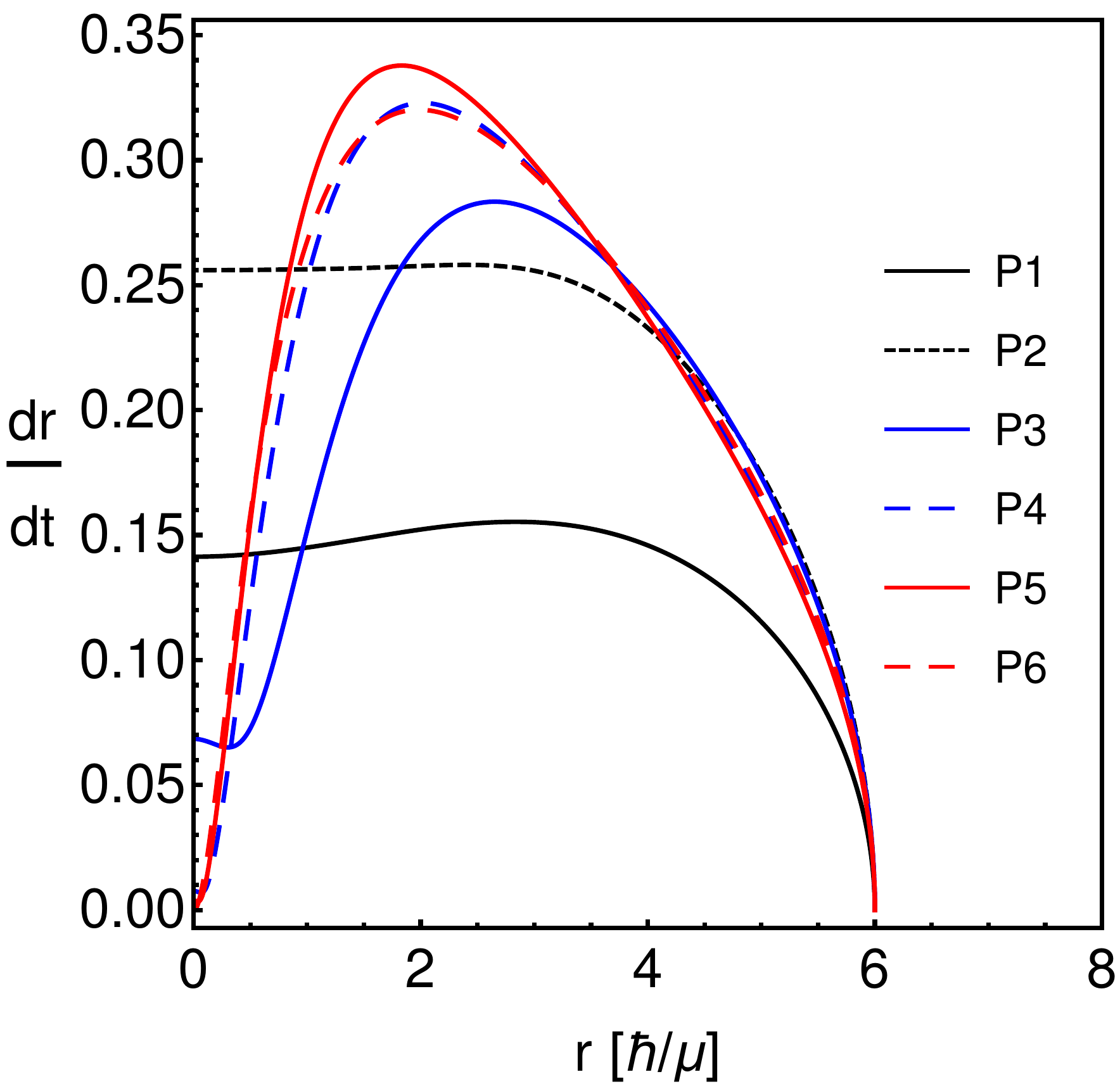}
\caption{Plot of $\frac{dr}{dt}$ as a function of $r$ for the orbits in Fig.~\ref{eff_potential_m1_j0}.
}
\label{figuredrdt-p}
\end{figure}

\begin{figure}[!htb]
	\includegraphics[width=\linewidth]{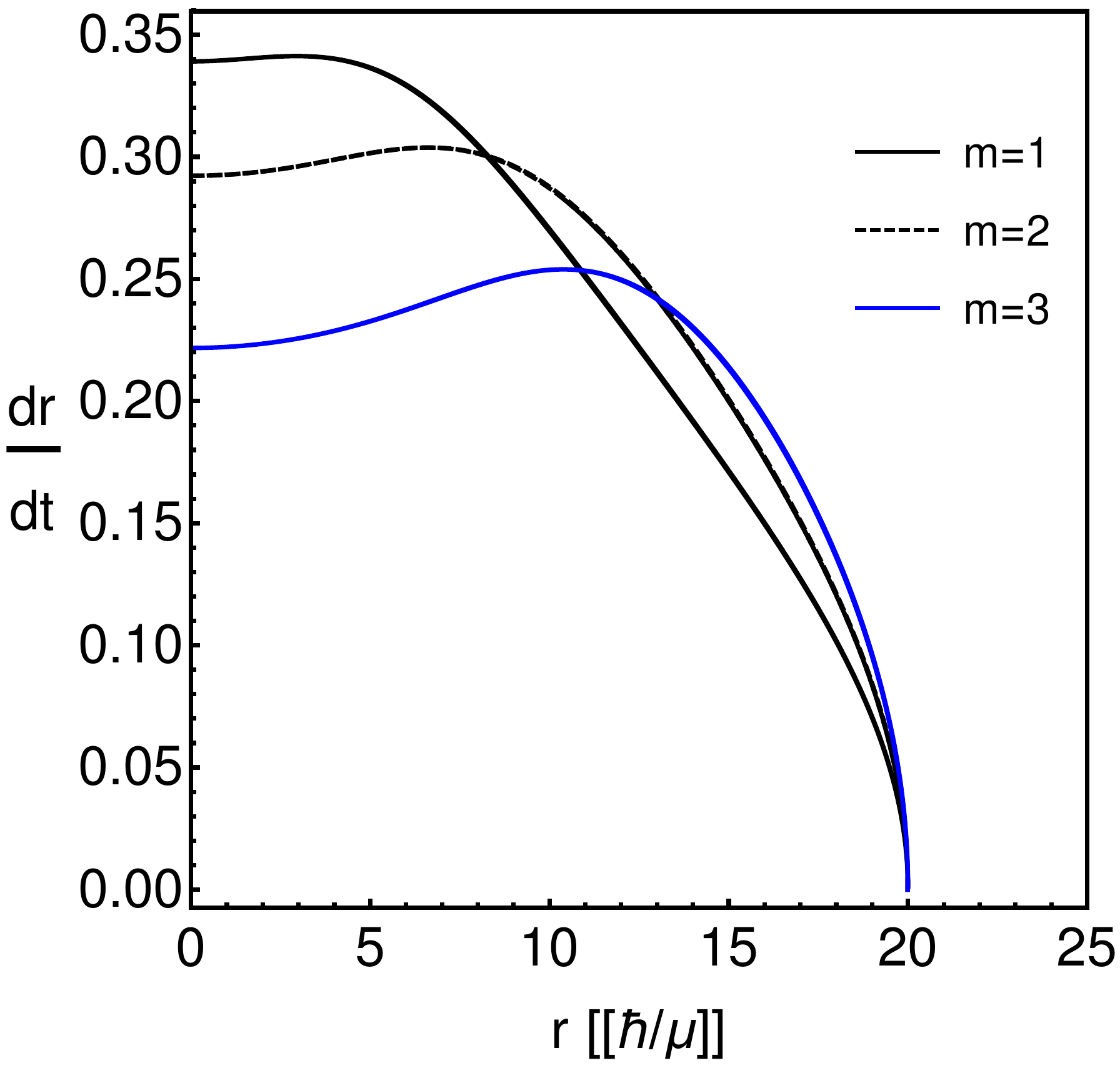}
	\caption{Plot of $\frac{dr}{dt}$ as a function of $r$ for the orbits in Fig.~\ref{orbits_m123_j0}.}
	\label{figuredrdt-p-2}
\end{figure}

Finally, we will quantitatively analyze the properties of our results about the orbits. By using
\begin{equation}
\frac{dr}{dt}=\frac{dr/d\tau}{dt/d\tau}=\frac{u^r}{u^t},
\end{equation}
we give the corresponding relation between the radial velocity and radial position of each orbit. We give respectively the results of $\frac{dr}{dt}$ as a function of $r$ in Figs.~\ref{figuredrdt-p} and \ref{figuredrdt-p-2} for the orbits described in Figs.~\ref{eff_potential_m1_j0} and \ref{orbits_m123_j0}. From the results in Fig.~\ref{figuredrdt-p}, we can see that the radial velocity $du/dt$ in the central region will approach to zero with the boson star becomes more and more compact, which means that a particle could stay for a long time in the central region of the rotating boson star.

\begin{figure}[!htb]
\includegraphics[width=\linewidth]{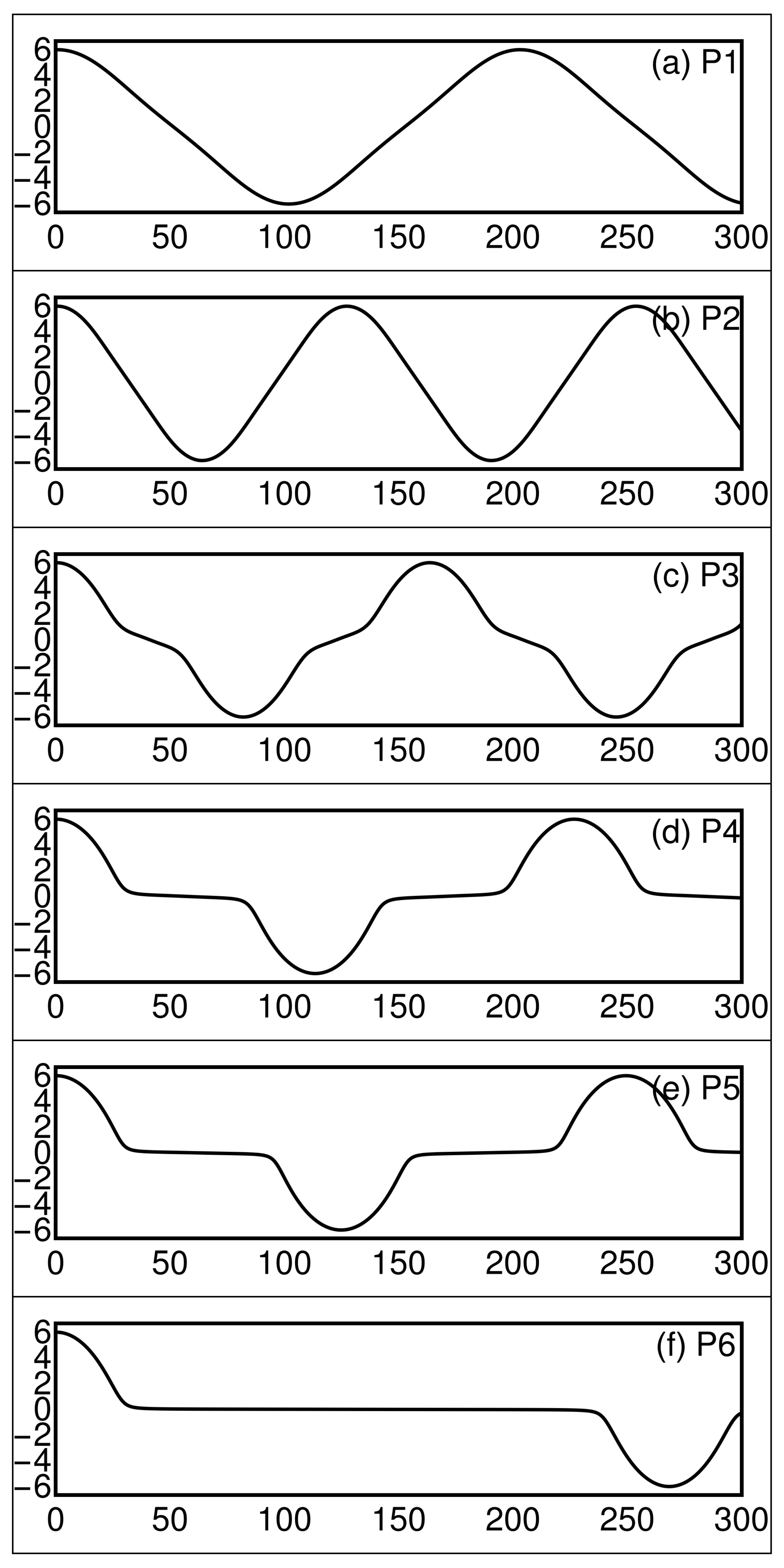}
\caption{Relations between the radius and coordinate time for the orbits in Fig.~\ref{eff_potential_m1_j0}, where the horizontal axis is the coordinate time and the vertical axis is the radius.}
\label{figurert-p}
\end{figure}

To measure the time that how long the test particle could stay in the central region of the rotating boson star, we give the results of the radius as a function of the coordinate time in Fig.~\ref{figurert-p} for the orbits described in Figs.~\ref{eff_potential_m1_j0}. It is easy to see that the time that the test particle stays in the central region of the rotating boson star increases with the boson star becomes more and more compact. The central region looks like a pocket to confine the test particle in the central region. If the rotating boson star is located in the center of a galaxy, such orbits will provide complete new type of observable effects.

The results of 16 years of monitoring stellar orbits around the Sagittarius A* \cite{Gillessen:2008qv} proved that there is a supermassive compact object with mass of $4\times 10^6 M_\odot$ locating at the center of the Milky Way, the information of the observed orbits around Sagittarius A* will be important for studying the properties and the central supermassive compact object. We take some part of the orbits in Ref. \cite{Gillessen:2008qv} and list their periapse values and orbital eccentricities in Table \ref{observed_orbits_sgra}. The maximal orbital eccentricity of the observed bound orbits is $e\sim 0.963$ and the corresponding periastron $r_p\sim 893.285M$. It is easy to see that the orbital angular momenta of these observed orbits are all much more larger than the orbital angular momenta of the orbits that we obtained in Figs. \ref{eff_potential_m1_j0}, \ref{eff_potential_m1_j005}, \ref{eff_potential_m2_j0}, \ref{eff_potential_m2_j005}, \ref{eff_potential_m3_j0}, and \ref{eff_potential_m3_j005}.

To obtain the possible deviation between orbits of the test particle in the backgrounds of the Kerr black hole {in the center of Sagittarius A*} and the rotating boson star. We should compare the orbits found in this paper to those around a black hole with the same mass and angular momentum. Note that, the dimensionless spin parameter $J/M^2$ of a rotating boson star will be greater than one and it is not acceptable for a Kerr black hole. Therefore, the orbits in the background of rotating boson star with the dimensionless spin parameter $J/M^2>1$ can not exist in the background of Kerr black hole. {However, the spin of the supermassive object in the center of Sagittarius A* has not been strictly determined by the observational data and it could be in the range of $a\in(0.1,0.98)$ \cite{Rockefeller:2005ta,Prescher:2005sd,Broderick:2008sp,Broderick:2010kx,Sanjeev2021}. The fact that there are no solid measurements of the spin of Sagittarius A* means we can not directly compare the  orbits in the backgrounds of the boson star and Sagittarius A* with an accurate spin. To show the differences of the orbits, we only consider the case of the rotating boson star with $k=1$ and $\omega=0.79$ and the Kerr black hole with the same ADM angular momentum and ADM mass. Here the corresponding ADM mass, ADM angular momentum, and the dimensionless spin parameter are taken as $M_{\text{ADM}}=1.3124$, $J_{\text{ADM}}=1.3776$, and $a=J/M^2=0.7998$.}

\begin{figure*}[!htb]
	\includegraphics[width=\linewidth]{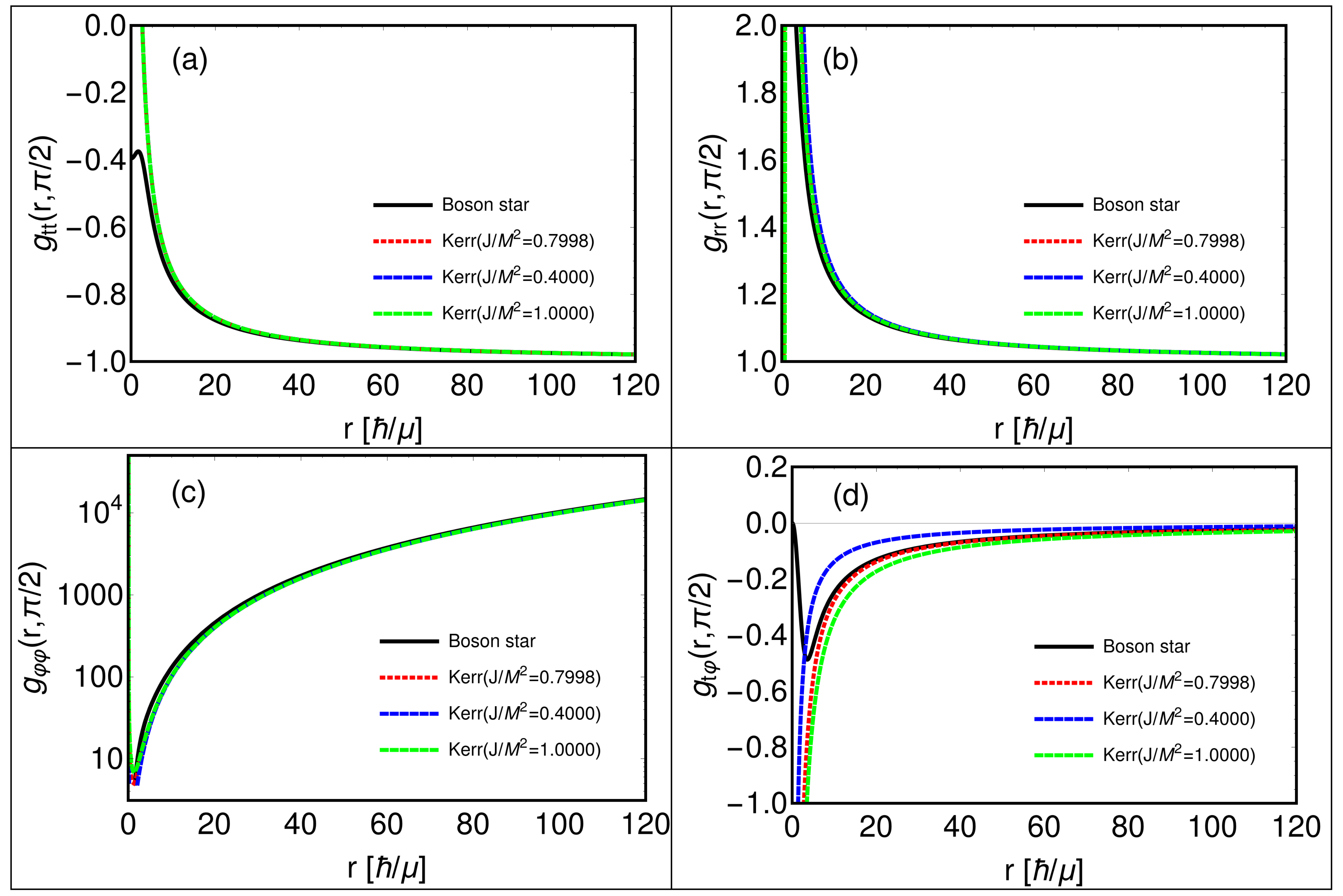}
	\caption{Relations of the metric functions between the rotating boson star with $(k=1, \omega=0.79)$ and the Kerr black hole with the same mass but different spins.}
	\label{com_rbs_kerr}
\end{figure*}

\begin{figure*}[!htb]
	\includegraphics[width=\linewidth]{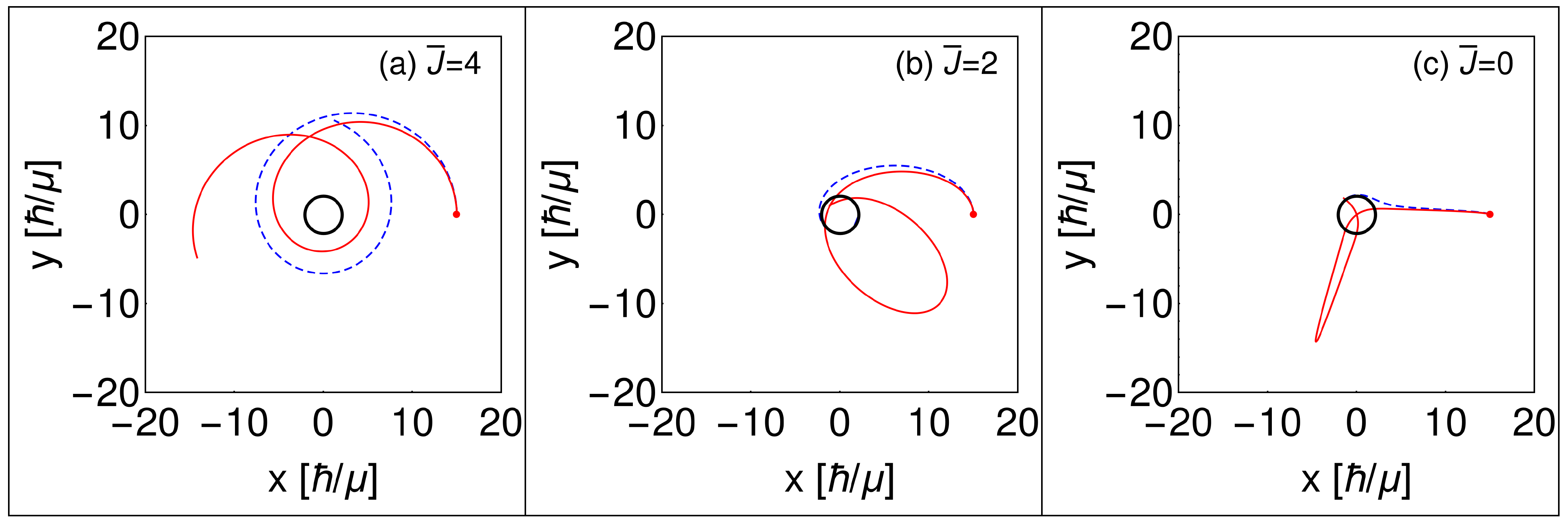}
	\caption{Orbits with the same orbital angular momenta and same initial positions ($r=15\hbar/\mu$) in the backgrounds of rotating boson star $(k=1, \omega=0.79)$ and Kerr black hole, where the masses and spins of the rotating boson star and Kerr black hole are the same, the black circle is the event horizon of the black hole.}
	\label{orbits_rbs_kerr}
\end{figure*}

We give the corresponding comparisons of the metric functions in the equatorial plane from both of them with different spin angular momenta in Fig. \ref{com_rbs_kerr}. It can be seen that, when the rotating boson star and the black hole have the same mass and spin angular momentum, their metric functions will tend to be consistent when the radius are much greater than the event horizon scale. The differences of the metric functions will increase when the radius $r$ decreases and the motion of test particle in the vicinity of the rotating boons star and Kerr black hole will be different even though the boson star and black hole have the same mass and spin angular momenta \cite{Grould12017}, see the results in Fig. \ref{orbits_rbs_kerr}.

Note that the minimal periastron is $r_p\sim 893.285M$ for the current observed stellar orbits around the Sagittarius A* \cite{Gillessen:2008qv}. Thus, the current observed stellar orbits do not have the enough accuracy to distinguish whether there is a rotating boson star or a Kerr black hole in the center of the Milky Way. We still give the comparisons for the metric functions of the rotating boson star and Kerr black holes with the same mass but different spin angular momenta in Fig. \ref{com_rbs_kerr}. It is shown that the difference of $g_{t\varphi}$ is larger than the others, which means one can use the information of the precession of the observed orbits to accurately study the nature of the Sagittarius A*. However, the current orbital observation data is not enough to accurately study the precession of the observed orbits.

\begin{table*}[!htb]
	
		\begin{center}
			\caption{Parameters of the observed stellar orbits around the Sagittarius A*, these data are obtained from the Table 7 in Ref. \cite{Gillessen:2008qv}.}
			\begin{tabular}{ c| c| c| c|| c| c| c| c }
				\hline
				\hline
				~~star~~&~~~~$r_p/M$~~~~&~~~~$r_a/M$~~~~&~~~~~$e$~~~~~&~~star~~&~~~~$r_p/M$~~~~&~~~$r_a/M$~~~&~~~~$e$~~~~\\
				\hline
				s1      &  24145.9      &	71671.0	    &   0.496     & s24    &  6697.75      & 193235.0    &	0.933\\
                s2      &  1391.99      &	21807.8	    &   0.880     & s27    &  4338.92      & 35836.3     &	0.784\\
                s4      &  16693.6      &	39513.9	    &   0.406     & s29    &  2055.16      & 83576.5     &	0.952\\
                s5      &  3725.17      & 43428.8	    &   0.842     & s31    &  3144.98      & 71735.6     &	0.916\\
                s6      &  4687.48      & 77549.1	    &   0.886     & s33    &  1854.85      & 54352.7     &	0.934\\
                s8      &  6821.86      & 70699.3	    &   0.824     & s38    &  10401.2      & 66931.3     &	0.731\\
                s9      &  4835.64      & 50428.8	    &   0.825     & s66    &  2595.54      & 23622.1     &	0.802\\
                s12     &  2904.69      & 55189.0	    &   0.900     & s67    &  93800.6      & 134425.0    &	0.178\\
                s13     &  14284.8      & 41734.1       &	0.490     & s71    &  65264.9      & 141270.0    &	0.368\\
                s14     &  893.285      & 47392.4       &	0.963     & s83    &  15609.5      & 184512.0    &	0.844\\
                s17     &  18653.7      & 40005.8       &	0.364     & s87    &  90088.2      & 435207.0    &	0.657\\
                s18     &  6022.98      & 43960.2       &	0.759     & s96    &  126618.      & 164793.	 &  0.131\\
                s19     &  4338.92      & 35836.3       &   0.784     & s97    &  143898.      & 268417.	 &  0.302\\
                s21     &  11740.2      & 138775.0      &	0.844     & s111   &  -            & -           &  1.105\\

				\hline
				\hline
			\end{tabular}
			\label{observed_orbits_sgra}
		\end{center}
\end{table*}

Here we give a simple summary about our results as follows:

(a) Due to the rotating boson star does not have a horizon and singularity, which means a test particle could pass through the center of the rotating boson star with a suitable energy and angular momentum. We check that for a test particle with a zero angular momentum, the value of the effective potential (\ref{effective-potential-r}) is finite and the test particle could pass through the center of the rotating boson star. When the angular momentum of the test particle is nonzero, the value of the effective potential in the center will be infinite and the test particle can not pass through the center of the rotating boson star.

(b) For a given boson star with a fixed angular momentum $m$, it will transit from the low rotating state to the highly relativistic rapidly rotating state with the increase of the compactness of the boson star. The background scalar field still becomes more and more compact. For a given boson star with fixed frequency $\omega$, the compactness of the background scalar field decreases with the angular momentum $m$. While the Lense-Thirring effect increases with $m$.

(c) When the boson star transforms from a low rotating state to a highly relativistic rapidly rotating state, the Lense-Thirring effect of the background on the orbit will be more and more significant. We find that the time of a test particle stays in the central region of the rotating boson star increases with the increase of the compactness of the boson star. Especially for the case (f) described in Figs. \ref{eff_potential_m1_j0} and \ref{eff_potential_m1_j005}. This is the pronounced difference for the small angular momentum orbits between a black hole or a rotating boson star, see the results given in Fig. \ref{figurert-p}. Here, we naively name this behavior as ``pseudo-plunge" into a rotating boson star.

(d) For the rotating boson star and black hole that have the same mass and spin angular momentum, the corresponding information of the current observed orbits around the Sagittarius A* can not accurately distinguish them. To observe the potential hints for distinguishing the black hole or boson star, we need to consider the closer orbit to the center of the star and the observation of orbits for a longer period of time.

\section{Conclusions and outlook}\label{Conclusion}

In this paper, we investigated the motion of a test particle in the equatorial plane of a rotating boson star with angular number $m=(1, 2, 3)$. We considered the boson star from low rotating state to highly relativistic rapidly rotating state. We solved the four-velocity of the test particle and derived the corresponding radial effective potential. The zero and nonzero angular momentum orbits of the test particle were derived. We still gave the periapse values of the orbits that we obtained, it was found that the bonson star could possess the orbits with orbital eccentricity $e=1$ and with small apastrons. With the rotating boson star changes from low rotating state to highly relativistic rapidly rotating state, it will have an ergoregion and the corresponding Lense-Thirring effect of the rotating boson star leads to the novel orbits described by Figs. \ref{eff_potential_m1_j0}, \ref{eff_potential_m1_j005}, \ref{eff_potential_m2_j0}, \ref{eff_potential_m2_j005}, \ref{eff_potential_m3_j0}, and \ref{eff_potential_m3_j005}.

We found that when the test particle moves in the rotating boson star with an ergoregion, it will stay in the central region of the boson star for a longer time. Especially for a highly relativistic rapidly rotating boson star, the Lense-Thirring effect on the motion of the test particle will be significant. Our results about the orbits and the corresponding profiles of $\frac{dr}{dt}$ along the radial direction show that the test particle could possess a trajectory that gets trapped for some time in the central region of the rotating boson. Such novel orbits will promote our further understanding of a rotating boson star and they will provide complete new type of observable effects of gravitational waves.

We compared the periapse values of the orbits that we obtained with the observed orbits around Sagittarius A* \cite{Gillessen:2008qv}. We compared the behaviors of the metric functions between the rotating boson star and Kerr black hole with the same mass and spin angular momentum and showed that they will tend to be consistent when the radius is of about $10^3M$. Note that for the current observed orbits around Sagittarius A*, the minimal periastron is of about $r_p\sim 893.285M$. To distinguish whether there is a rotating boson star or a Kerr black hole in terms of the orbits, the orbits with small periapse values are necessary. We also gave the orbits of test particle with different orbital angular momenta in the backgrounds of the rotating boson star and Kerr black hole with the same mass and angular momentum in Fig. \ref{orbits_rbs_kerr}. It was found that the orbits with small orbital angular will plunge into the black hole or pass through the central region of the boson star. Obviously, it is easy to distinguish the boson star and black hole with such orbits. However, the information of the current observed orbits around Sagittarius A* can not give us enough information to recognize whether Sagittarius A* is a black hole or a boson star.

We should note that the existence of an ergoregion for a rotating boson means that such system is unstable and the final state of an unstable rotating boson star should be investigated by using the non-linear evolution \cite{Siemonsen:2020hcg}. Recently, the stable multi-state rotating boson stars have been proposed in Refs. \cite{Li2020, Li2020ffy, Sanchis-Gual:2021edp}. It was shown that the single field rotating boson star can also be stabilized using nonlinear scalar self-interactions \cite{Siemonsen:2020hcg}. But the horizonless properties still make sure that they still have the similar kinds of orbits that we obtained in this paper. Therefore our results are still useful for understanding the properties of boson stars.


\section{Acknowledgments}

We thank Prof. Philippe Grandcl\'ement for the help about using the spectral solver KADATH. This work was supported in part by the National Key Research and Development Program of China (Grant No. 2020YFC2201503), the National Natural Science Foundation of China (Grants No. 12105126, No. 11875151, No. 12075103, and No. 12047501), the Fundamental Research Funds for the Central Universities (Grant No. lzujbky-2021-pd08), the China Postdoctoral Science Foundation (Grant No. 2021M701531), and the 111 Project (Grant No. B20063). Y.X. Liu was supported by Lanzhou City's scientific research funding subsidy to Lanzhou University.

\end{document}